%% file: main.tex
\documentclass{article}

\usepackage{PRIMEarxiv}

\usepackage[utf8]{inputenc} 
\usepackage[T1]{fontenc}    
\usepackage{url}            
\usepackage{booktabs}       
\usepackage{amsfonts}       
\usepackage{nicefrac}       
\usepackage{microtype}
\usepackage{graphicx}       
\usepackage{natbib} 

\usepackage{setspace} 

\usepackage{float}
\usepackage{ragged2e}
\usepackage{caption}
\usepackage[most]{tcolorbox}
\usepackage{listings}
\tcbuselibrary{breakable, skins}

\usepackage{tikz}
\usetikzlibrary{shapes.geometric, arrows.meta, positioning, calc}

\usepackage{chngcntr}

\newcommand{\setappendixfigures}[1]{
  \renewcommand{\thefigure}{#1.\arabic{figure}}
  \setcounter{figure}{0}
}

\title{Multimodal LLMs for Historical Dataset Construction from Archival Image Scans: \\ German Patents (1877--1918)\thanks{Paper website: \url{https://historymind.ai}}\hspace{0.4em}\thanks{Code: \url{https://github.com/niclasgriesshaber/llm_patent_pipeline.git}}\hspace{0.4em}\thanks{Archival Image Corpus: \url{https://digi.bib.uni-mannheim.de/sammlungen/patentregister}}}

\author{
  Niclas Griesshaber\thanks{Corresponding author. Now at the University of Oxford.} \\
  University of Mannheim \\
  \texttt{niclas.griesshaber@history.ox.ac.uk} \\
   \And
  Jochen Streb \\
  University of Mannheim \\
  \texttt{jochen.streb@uni-mannheim.de} \\
}

\usepackage{hyperref}       
\hypersetup{
    colorlinks = true,
    citecolor = black,
    linkcolor = black,
    urlcolor  = blue,
    pdfborder = 0 0 0
}

\newcommand{\AppendixA}{\hyperref[sec:appendix_a]{Appendix A}}
\newcommand{\AppendixB}{\hyperref[sec:appendix_b]{Appendix B}}

\usetikzlibrary{shapes.geometric, arrows.meta, positioning, calc, decorations.pathreplacing}

\begin{document}

\maketitle

\setcounter{footnote}{0}

\input{00_abstract}
\input{01_introduction}
\input{02_historical_background}
\input{03_related_work}
\input{04_data}
\input{05_methodology}
\input{06_benchmarking}
\input{07_economics_of_llms}
\input{08_conclusion}

\section*{Acknowledgments}
We are grateful for feedback from Philipp Ager, Stephen Broadberry, Alexander Donges, Gavin Greif, Leander Heldring, John Meyer, Sheilagh Ogilvie, the audience at the Mannheim PhD colloquium in September 2025, the audience at the launch event of the Oxford Digital Scholarship Society in November 2025, the audience at the Oxford Graduate Seminar in November 2025, and participants of the Oxford-Warwick-LSE (OWL) workshop in December 2025. Furthermore, we are highly thankful to Carsten Burhop, who lent us a large share of the physical volumes. We also thank Tünde Gottschling and Irene Schumm for their unwavering support in managing the scanning of all primary sources in the Digital Lab of the University of Mannheim. Finally, we extend thanks to our research assistants, in particular Julius Schöffel, for scanning the physical sources, constructing the benchmarking datasets, and manually cleaning the patent dataset. Our research was financially supported by the Heidelberg Academy of Sciences as part of its academy program ``Finanz- und Unternehmensforschung in Langfristperspektive''. Niclas Griesshaber is highly grateful for the DPhil funding provided by the Economic and Social Research Council through UK Research and Innovation under the Advanced Quantitative Methods Award.

\section*{AI Transparency Statement}
The first draft of this paper was written by the authors without the assistance of any external tools. Afterward, LLMs were used to restructure logical flows, correct grammar and selectively deployed to improve readability by suggesting synonyms and alternative phrasing using the wording from our first draft. The code for our GitHub repository and the project website was generated completely by LLMs. In particular, we used the Cursor code editor, where we guided the coding agents with natural language to build our website and pipeline. We reviewed the LLM-generated code to the best of our ability to ensure a responsible use of AI when creating our codebase and website.

\newpage
\bibliographystyle{agsm-custom}
\bibliography{references} 

\newpage
\input{09_appendix_a}
\input{10_appendix_b}

\end{document}

%% file: 00_abstract.tex
\begin{abstract}
We leverage multimodal large language models (LLMs) to construct a dataset of 306,070 German patents (1877--1918) from 9,562 archival image scans using our LLM-based pipeline powered by Gemini-2.5-Pro and Gemini-2.5-Flash-Lite. Our benchmarking exercise provides tentative evidence that multimodal LLMs can create higher quality datasets than our research assistants, while also being more than 795 times faster and 205 times cheaper in constructing the patent dataset from our image corpus. About 20 to 50 patent entries are embedded on each page, arranged in a double-column format and printed in Gothic and Roman fonts. The font and layout complexity of our primary source material suggests to us that multimodal LLMs are a paradigm shift in how datasets are constructed in economic history. We open-source our benchmarking and patent datasets as well as our LLM-based data pipeline, which can be easily adapted to other image corpora using LLM-assisted coding tools, lowering the barriers for less technical researchers. Finally, we explain the economics of deploying LLMs for historical dataset construction and conclude by speculating on the potential implications for the field of economic history.
\end{abstract}

\keywords{Multimodal Large Language Models \and Information Extraction \and Dataset Construction \and German Patents}

%% file: 01_introduction.tex
\section{Introduction}
\label{sec:01_introduction}

For a long time, economic historians focused on compiling and interpreting long macroeconomic time series, with GDP per capita playing the most prominent role \citep{maddison2006development,broadberry2015british}. This inevitably led to methodological limitations, as microeconomic data is needed to understand how consumers and producers react to economic policy changes or exogenous shocks. In particular, it is important to consider the heterogeneity of historical actors. People differ in terms of age, gender, education, social status, income, and wealth, among other things, and depending on these differences, they make different decisions under the same external conditions. In order to take this heterogeneity into account, large-scale microeconomic datasets are required. 

The fact that economic historians rarely use microeconomic data are not due to their inaccessibility. Socioeconomic data on consumers can be obtained, for example, from historical census data \citep{long2013intergenerational,ruggles2014big, abramitzky2021automated}, savings books \citep{lehmann2018does}, information on companies from patent statistics \citep{DongesStreb+2024+301+323}, stock market newspapers \citep{lehmann2024data}, city directories \citep{albers2023perks}, or trade registers \citep{guinnane2007putting}. However, accessing these data was associated with high and sometimes prohibitive costs because the manual construction of the dataset by research assistants was slow, expensive, and error-prone. To accelerate historical dataset construction, methods from the digital humanities and computer science, such as optical character recognition (OCR), named-entity recognition (NER), or information extraction (IE), have become increasingly attractive within the field of economic history \citep{shen2020large,dell2023american,bergeaud2024new}. However, these digital methods require advanced programming and labeled data to build custom pipelines that work only on homogeneous image corpora. Recent advances in artificial intelligence (AI) have the potential to remove these barriers and make large-scale dataset construction accessible to non-technical researchers, thereby eliminating the field's major bottleneck: transforming archival image scans into a structured dataset.

Multimodal large language models (LLMs) are neural networks that can jointly process multiple types of data. For example, in this paper we send images and raw text instructions to multimodal LLMs to construct a large-scale historical patent dataset. Multimodal LLMs have shown remarkable performance across a wide range of tasks, including language and vision \citep{devlin2019bert,brown2020language,dosovitskiy2020image}. They are---like many of the existing methods from OCR, NER, and IE---based on the deep learning paradigm, which refers to neural networks with many layers that can learn complex patterns from data \citep{lecun2015deep,goodfellow2016deep}. When LLMs are scaled with data, parameters, and computing capacity, they become surprisingly powerful \citep{kaplan2020scaling,hoffmann2022training, wei2022emergent}, offering a unified approach to OCR, NER, and IE. These immense economies of scale are also the reason why the most capable LLMs are almost exclusively developed by private AI companies.

The advancements and commercialization of multimodal large language models have enabled us to make several contributions. First, to construct a dataset that contains all 306,070 patent entries that are included in 41 physical volumes of historical patent registers published between 1877 and 1918, we used LLM-assisted coding tools to develop an LLM-based data pipeline that processed our 9,562 image scans. We then extracted five variables from each patent entry: \textit{patent\_id}, \textit{assignee}, \textit{location}, \textit{title}, and \textit{date}. Moreover, our research assistants manually created a \textit{student-constructed} benchmarking dataset from 41 randomly selected images, one from each of the annual patent registers used. Initially, we compared the \textit{LLM-generated} dataset against the \textit{student-constructed} dataset to evaluate the reliability of our LLM-based pipeline---especially to check for hallucinations that would undermine the data quality and bias any subsequent econometric analysis. Surprisingly, we realized that many highlighted differences between both datasets were not due to errors made by the LLM, but due to mistakes made by our research assistants. For this reason, we also constructed a second, \textit{perfect} benchmarking dataset. By comparing the \textit{student-constructed} and \textit{LLM-generated} dataset to the \textit{perfect} dataset, we provide tentative evidence that multimodal LLMs are able to produce higher quality datasets from our image corpus than our research assistants. Finally, we open-source both our LLM-based data pipeline and the new historical patent dataset to accelerate empirical research in economic history.

This paper is structured as follows: Section~\ref{sec:02_historical_background} provides the historical background of our data. Section~\ref{sec:03_related_work} then reviews related work on the construction of patent databases, including manual work, custom machine learning approaches, and the emerging literature on LLMs in economic history. Section~\ref{sec:04_data} describes our primary sources that form our archival image corpus. Section~\ref{sec:05_methodology} presents our LLM-based data pipeline to construct the patent dataset, followed by Section~\ref{sec:06_benchmarking} which addresses data quality and hallucination concerns using our benchmarking datasets. Section~\ref{sec:07_economics_of_llms} explains the economics of constructing datasets from image scans using multimodal LLMs. Section~\ref{sec:08_conclusion} concludes and discusses the potential implications of multimodal LLMs for the field of economic history.

%% file: 02_historical_background.tex
\section{Historical Background}
\label{sec:02_historical_background}

Modern economic growth, which began in Great Britain in the early nineteenth century and has since been emulated by many countries around the world, is based on a never-ending stream of innovation. A central question in economic history is therefore why it took so long for innovations to evolve from a rare occurrence to a constant driver of growth \citep{galor2011unified, mokyr2011gifts}. To understand which factors promote and hinder the emergence of innovation, it is first necessary to measure the type and scope of innovations as precisely as possible. In economic history research, this is done primarily on the basis of patent statistics.\\

The origins of the patent system date back at least to mid-fifteenth century Venice. However, it was not until the end of the eighteenth century that patents evolved from a privilege into an intellectual property right that could be obtained by registration with a patent authority. All patent authorities charged a filing fee for granting a patent; some also examined the novelty of a registered invention as an additional requirement. Since then, a patent holder has had the exclusive right to manufacture and market the protected innovation within the geographical scope of the patent law and for a limited period of time. For patent protection to work, potential imitators must be informed about which innovations are protected. Patent authorities have therefore long provided information about patents granted and have also collected detailed patent descriptions so that interested parties can consult them if necessary \citep{moser2011,Cox2019}. These historical sources are still available today and allow databases to be created with microeconomic information about the inventor and co-inventors, their gender and place of residence, as well as the number, title, and technological class of the patent.\\

The comparative ease of access to this mass of data is the great advantage of patents as a measure of the extent and direction of technological progress. Critics of this approach like to point out that not all innovations are patented, either because patent law does not allow them to be or because inventors believe that innovation gains are more likely to be realized through secrecy. Added to this is the problem that pure patent counts neglect the fact that only a minority of patents granted protect valuable innovations, while most patents represent rather worthless ideas that hardly constitute incremental innovations \citep{605a1bfd-4d7e-3033-86a6-6dbe02a35f0f,doi:10.1086/663631}. The first criticism concerns an undeniable weakness of patent statistics as a measure of innovation, which has not yet been remedied but is accepted by most economic historians. The alternative method proposed by Petra Moser, which is to identify historical innovations with the help of world exhibition data, has its own shortcomings \citep{moser2005patent,domini2020exhibitions}. The second criticism has since lost importance because economic historians have found various ways to assess the quality of patents \citep{streb2024cliometric}.\\

%% file: 03_related_work.tex
\section{Related Work}
\label{sec:03_related_work}

\subsection{Manual Dataset Construction}

Constructing datasets from primary sources by hand is very labor-intensive. Over the past decades, a large number of historical patent databases have been created manually. These databases now exist for the early industrialized countries of Great Britain, France, the United States, and the Netherlands, for the Mediterranean countries of Italy and Spain, for the Scandinavian countries of Finland, Norway, and Sweden, for Japan and even for some Latin American countries such as Argentina, Cuba, and Mexico \citep{streb2023patent}. 

In the mid-2000s, \citet{streb2006technological} decided to create a historical patent database for the German Empire and the Weimar Republic. However, they did not have sufficient financial resources to pay all the research assistants who would have been needed to manually transfer the more than half a million patents granted in Germany between 1877 and 1932 into a digital database. As a result, the economic historians had to forgo completeness and restricted their database to a subset that only included the most valuable patents. Because the German patent system levied an annual patent renewal fee, which was intended to encourage patent holders to abandon unprofitable patents as quickly as possible rather than retaining them for the maximum term of 15 years, it is possible to infer the economic value of a patent from its life span. They selected patents for the database that had been held for at least ten years from the date of grant. The resulting database comprises around 68,700 long-lived patents, which corresponds to about 10\% of all patents granted in the German Empire and the Weimar Republic. Despite the considerable reduction in the amount of data, it took several months to manually extract all the relevant information, perfectly illustrating the scalability limits of manual dataset construction.\\

\subsection{Custom OCR and NER Pipelines}

To overcome the constraints of manual dataset construction, recent work has utilized narrow task-specific models from machine learning. With regard to the creation of patent databases, ``PatentCity'' represents the state-of-the-art approach in this paradigm. \citet{bergeaud2024new} use traditional OCR algorithms and train their own deep learning models to construct a patent database. Due to this transition from manual to digital methods, the amount of data they were able to collect exceeds all previous ones. Their database contains information on the assignees and inventors of all patents granted in Germany between 1877 and 1980, in France between 1903 and 1980, in the United Kingdom between 1893 and 1980, and in the United States between 1836 and 1980. Where specified, the inventor's profession, citizenship, and address were also recorded. Furthermore, they assigned the inventors' addresses to a county or a municipality. 

To construct the PatentCity database, the authors first converted the historical patent specifications available as image scans into editable text files using Tesseract v5.0 and their own in-house OCR algorithm. In a second step, they trained a custom NER model to extract the relevant variables from the OCR text and applied an relationship algorithm to reconstruct how the entities interrelate. To train the NER model, the authors had to create a manually annotated dataset that was split into a training and test set. The former was used to teach the deep-learning model how to extract the desired entities. After the training had finished, they evaluated the model's performance by comparing its predictions with the ground truth on the test set. 

However, such narrow machine learning approaches face their own bottlenecks. The method by \cite{bergeaud2024new} is not universal and they had to retrain their NER model whenever the layout of the image scans changed because the structure of the OCR text shifted accordingly. This always required a new annotated dataset. Moreover, much of the spatial relationship of the embedded text on an image is lost when they perform OCR, which is crucial when constructing datasets from more complex historical documents depicting double-column layouts, tables, or maps. Finally, LLM-assisted coding tools did not exist when they first published their work, making it very difficult for non-technical researchers to adopt their pipeline to other image corpora.

The arrival of multimodal LLMs may offer a universal solution for historical documents as it mitigates all of the aforementioned issues, which enabled us to partially replicate the PatentCity database for German patents from 1877 to 1918 despite using a different primary source. Instead of patent specifications depicting information on a single patent per image scan, our source contains between 20 to 50 short patent entries per page, which are arranged in a double-column format. Furthermore, we go beyond replication by providing all patent titles that are not included in the PatentCity database.

\subsection{Large Language Models in Economic History}

There is an emerging literature on large language models for text-as-data approaches in economics and economic history \citep[e.g.,][]{bartik2025costs,chyn2025ideology,GriesshaberOgilvie2025,lagakos2025american,ash2025interviews}. Existing work uses LLMs to extract higher-level features from editable text to construct variables for downstream econometric analysis. For example, \citet{GriesshaberOgilvie2025} use GPT-4o to classify sentences into broader economic categories to examine institutional differences within the guild system across colonial Latin America. \citet{lagakos2025american} extract features from biographies to understand the sources of a meaningful life in early-twentieth-century America. In both cases, LLMs lead to unprecedented productivity increases as economic historians and research assistants do not have to read and manually classify the thousands of biographies and guild regulations. However, as LLMs were trained on a finite dataset, they inevitably make biased predictions, especially when those predictions involve value judgments. \citet{carlson2025unifying} show that this bias can propagate to downstream estimators. They further emphasize that researchers may use various LLMs with different biases to achieve significant results in their regression analyses. For this reason, the authors propose the MAR-S framework, which includes the construction of a validation dataset to correct for the bias introduced during the LLM's feature extraction.

Compared to these text-as-data approaches, value judgments are usually less pronounced when merely transcribing printed, Gothic or handwritten text from image scans. Multimodal LLMs can be given an image together with a text instruction, prompting them to transcribe the embedded text. There are several factors that influence the LLMs' transcription accuracy: font type, image size, information density, resolution, document degradation, prompt, and many more. As the immense heterogeneity of archival sources affects all of these factors, it is difficult to assess the general reliability of multimodal LLMs on text transcription from image scans. \citet{greif2025multimodal} benchmark the OCR capabilities of GPT-4o, Gemini-2.0-Flash, and Transkribus' Text Titan I on eighteenth- and nineteenth-century German city directories, printed in Roman and Gothic. They report the best transcription results for Gemini-2.0-Flash. \citet{humphries2025unlocking} find that LLMs transcribe eighteenth- and nineteenth-century English handwritten documents with higher accuracy than fine-tuned models by Transkribus. \citet{levchenko2025evaluatingllmshistoricaldocument} also reports this for Russian prints in Civil font, with Gemini-2.5-Pro yielding the best performance among 12 evaluated multimodal LLMs. \citet{crosilla2025benchmarking} show the superiority of LLMs over Transkribus on modern handwriting. Even more astonishing than the universal transcription capabilities of multimodal LLMs is the fact that we still do not fully understand how they accomplish this task. As a result, the nascent subfield of mechanistic interpretability tries to cast light on the internal mechanisms behind this ability \citep{baek2025largevisionlanguagemodelstext,kim2025interpretingattentionheadsimagetotext}.

Perhaps even more striking than the ability of multimodal LLMs to transcribe embedded text in images is that they can be prompted to extract only the desired information in order to construct a structured dataset. Intuitively, strong transcription capabilities are a prerequisite for this task. The economic historian Jonathan \citet{jayes2025like} employs Gemini-2.0-Flash to extract information from image scans of Swedish firm reports. As more and more economic historians use multimodal LLMs to create large-scale datasets, it will be necessary to benchmark multimodal LLMs rigorously on the task of information extraction, and by extension, on the task of historical dataset construction. \citet[p.~1]{xie2025multimodal} find that GPT-4o ``generates usable yet imperfect data'' when deployed to extract information from Swedish patent cards (1945--1975). \citet{luo2025multimodal} expands the literature on information extraction on Swedish patent cards by benchmarking several open-source and proprietary LLMs on the same sample, with Gemini-2.5-Pro achieving the strongest performance. Likewise, parallel work by \citet{vafaie2025end} evaluates GPT-4o-mini and open-source models like InternVL2.5 for key information extraction on a new annotated dataset from heterogeneous German index cards. However, the accuracy of all these models is insufficient for robust research in economic history. Moreover, a report on work in progress by \citet{rodrigues2025benchmarking} finds that Gemini-2.5-Flash outperforms GPT-4o when creating JSON entries for two-column printed bibliographies. Finally, \citet{backer2025can} developed an LLM-based pipeline to extract information from historical tables to build a panel dataset. They also manually construct a dataset and show that their downstream regression analyses are ``statistically indistinguishable whether using LLM or gold standard data'' \citep[p.~1]{backer2025can}. They claim that multimodal LLMs offer a ``watershed change for the digitization of historical tables'' \citep[p.~1]{backer2025can}, which is even more striking as they use Claude-3.5-Sonnet and Gemini-1.5-Pro---multimodal LLMs that are much weaker than the models we use in this paper.\footnote{Based on this work, fellow researchers at the Federal Reserve Bank of Philadelphia have released a contemporaneous report on the institutional potential of multimodal LLMs for historical data collection \citep{moulton2025harvesting}.}

%% file: 04_data.tex
\section{Data}
\label{sec:04_data}

\begin{figure}[htpb]
\vspace{-1.15cm}
\caption{\\\textsc{THE PRIMARY SOURCE}}
\label{fig:representative-page}

\centering
\includegraphics[width=0.9\textwidth]{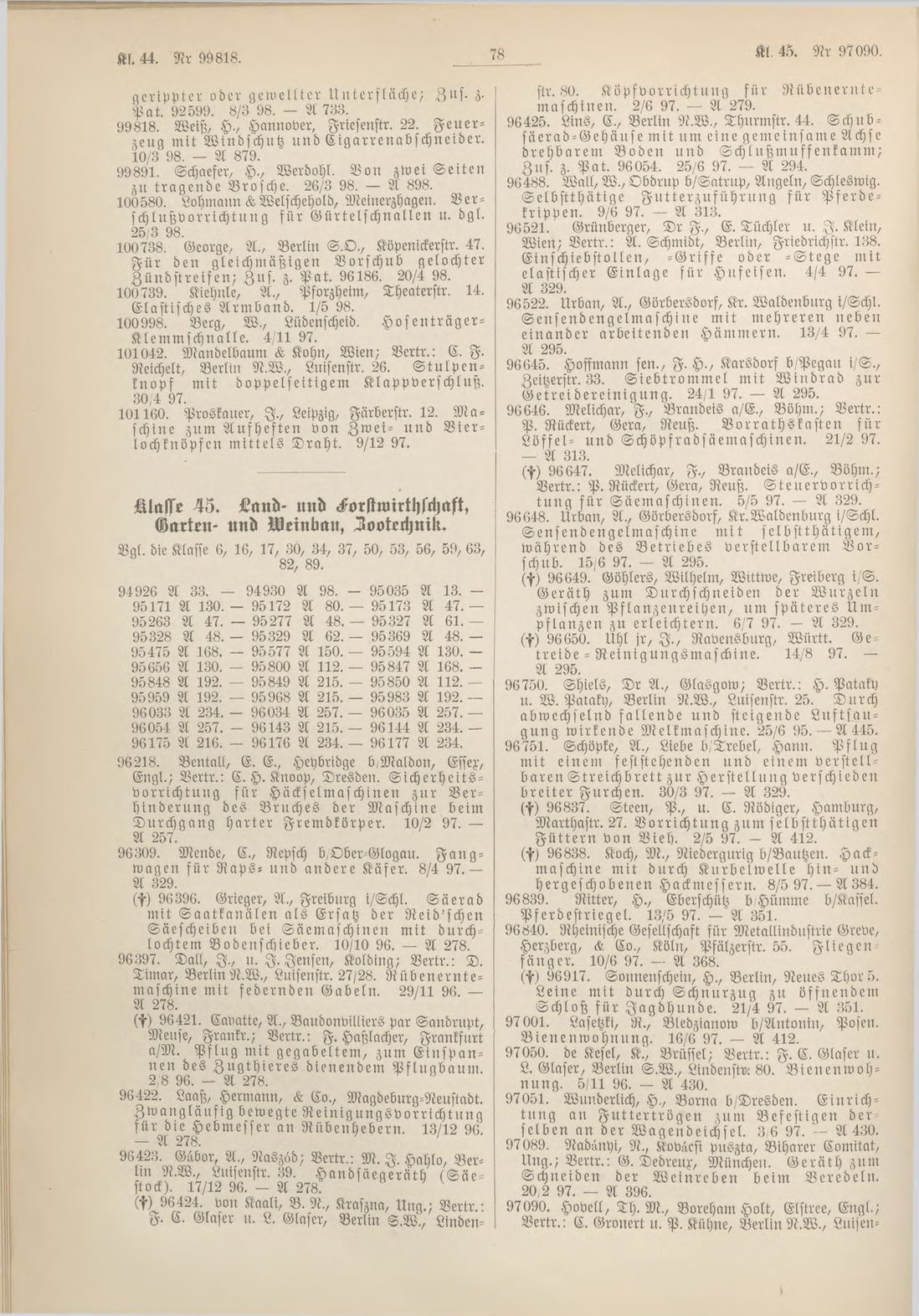}

\captionsetup{justification=raggedright,singlelinecheck=false}

\vspace{0.25cm}
\begin{minipage}{0.9\linewidth}
    \footnotesize
    \justifying
    \begin{spacing}{1}
    \noindent\textit{Notes}: A representative page from our 9,562 image scans (volume 1898, page 73). Please note how the patent entries at the top left and bottom right span multiple pages, as well as the single patent entry spanning both columns. Also note how much information we do not want to include, such as the running footer or the list of patent IDs that appears after the introduction of a new technological class. The introduction of a new technological class occurs randomly across both content columns. Moreover, many pages do not even introduce a new technological class.
    \end{spacing}
\end{minipage}

\end{figure}

\begin{figure}[htbp]
\centering

\caption{\\\textsc{A PATENT ENTRY}}
\label{fig:patent-entry}

\includegraphics[width=\textwidth]{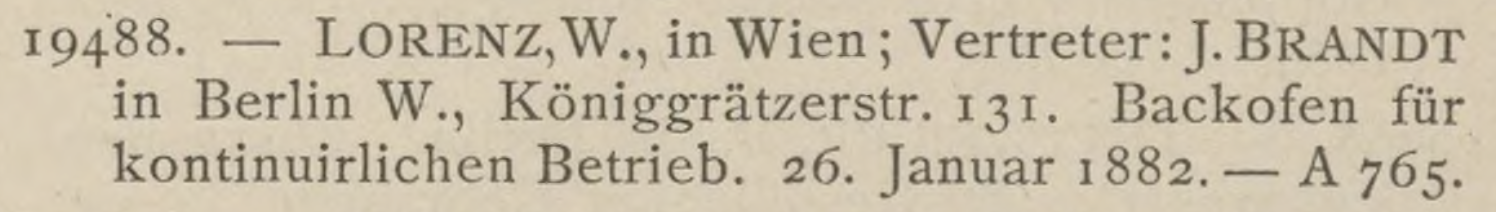}

\vspace{0.25cm}
\begin{minipage}{\linewidth}
    \footnotesize
    \justifying
    \begin{spacing}{1}
    \noindent\textit{Notes}: A representative patent entry depicted on page 1 in volume 1882. This patent lists a legal representative because it was registered by an Austrian. It does not include a priority claim, which would typically appear at the very end.
    \end{spacing}
\end{minipage}

\end{figure}

In Germany, a nationwide patent law was not introduced until 1877, replacing the numerous independent patent laws of the German states that had previously been in force \citep{donges2019legal}. To provide the interested public with an overview of the newly granted patents, the Reich Patent Office published an annual volume listing all patents entered into the register in the previous calendar year. This annual patent register did not list the newly granted patents in numerical order. The order of the patents was based on the 89 different technological classes used by the Patent Office. First, all patents in class 1, ``Preparation of ores'', were listed in numerical order, followed by all patents in class 2, ``Baking'' (for an example, see page 312 in \citealp{DongesStreb+2024+301+323}), and the list always ended with patents in technological class 89, ``Sugar production''. Figure~\ref{fig:representative-page} depicts a representative page of our corpus. Several pieces of information were provided for each of these patents: the patent number, the name and address of the patent holder and, in the case of foreign patent holders, the name and address of their legal representative in Germany, the title of the patent, and the date of application, to which a priority claim is appended in rare cases (Figure~\ref{fig:patent-entry}). 

We were able to track down all volumes that were issued by the Reich Patent Office from 1877 until 1918. In total, we collected 41 physical books that were scanned by the Research Data Center at the University of Mannheim at a resolution of 300 dpi per image. A completeness check was conducted to ensure that not a single page was inadvertently skipped after which the TIFF images were merged into a PDF. For our dataset, we were only interested in each volume's chapter ``Systematische Übersicht'' that provides the list of newly granted patents described above. The layout of this chapter is fairly consistent across all 41 volumes. Patent entries are always short paragraphs, which are arranged in this double-column format. Likewise, the beginning of each of the 89 technological classes is introduced as a heading within one of the two main content columns. Finally, every page has a running header with the current patent number and technological class.

\newpage

However, there are also several fine-grained differences across our image corpus. Most notably, the general font type changes from Roman to Gothic in 1894. Within the Gothic period, the volumes from 1894 to 1911 use the Unger typeface, while those from 1911 to 1918 are set in Breitkopf. Later volumes introduce technological subclasses (e.g., ``17a'', ``89k'') as the number of patent registrations increased. The classifications become even more granular as patent numbers are preceded by enumerations (e.g., ``17. 287909'') to indicate the patent group. In addition, there are many minor variations across volumes. For example, the format of the registration date varies and some patent numbers are preceded or followed by a dagger symbol ``(†)'', denoting patents that had already expired before the volume was published.

%% file: 05_methodology.tex
\section{Multimodal LLMs for Dataset Construction from Image Scans}
\label{sec:05_methodology}

\begin{figure}[htbp]
\vspace{2cm}

\caption{\\\textsc{OUR LLM-BASED DATA PIPELINE}}
\label{fig:pipeline-overview}

\centering

\resizebox{0.85\textwidth}{!}{
    \begin{tikzpicture}[
        node distance = 0.9cm and 1.5cm,
        process/.style = {
            rectangle, 
            draw=black, 
            semithick, 
            align=center, 
            inner sep=8pt,
            minimum width=4.2cm, 
            font=\small
        },
        prompt/.style = {
            rectangle, 
            draw=black, 
            dashed, 
            semithick, 
            align=center, 
            inner sep=6pt,
            font=\small\itshape,
            minimum width=3.5cm
        },
        arrow/.style = {
            -{Stealth[length=3mm, width=2mm]},
            thick
        },
        lbl/.style = {
            font=\footnotesize,
            midway,
            right=3pt,
            align=left
        }
    ]

        \node [process] (corpus) {Image Corpus \\ from a specific Volume};
        \node [process, below=1.3cm of corpus] (gemini_pro) {Gemini-2.5-Pro};
        \node [process, below=1.0cm of gemini_pro] (trunc_data) {Dataset with truncated\\patent entries};
        \node [process, below=1.3cm of trunc_data] (gemini_flash1) {Gemini-2.5-Flash-Lite};
        \node [process, below=1.0cm of gemini_flash1] (repaired_data) {Dataset with repaired\\patent entries};
        \node [process, below=1.3cm of repaired_data] (gemini_flash2) {Gemini-2.5-Flash-Lite};
        \node [process, below=1.0cm of gemini_flash2] (final) {LLM-generated Dataset};

        \node [prompt, left=of gemini_pro] (prompt1) {Patent Entry\\Extraction Prompt};
        \node [prompt, left=of gemini_flash1] (prompt2) {Reparation Prompt};
        \node [prompt, left=of gemini_flash2] (prompt3) {Variable Extraction\\Prompt};

        \draw [arrow] (corpus) -- (gemini_pro) node [lbl] {For each image};
        \draw [arrow] (gemini_pro) -- (trunc_data);
        \draw [arrow] (trunc_data) -- (gemini_flash1) node [lbl] {For each row};
        \draw [arrow] (gemini_flash1) -- (repaired_data);
        \draw [arrow] (repaired_data) -- (gemini_flash2) node [lbl] {For each row};
        \draw [arrow] (gemini_flash2) -- (final);

        \draw [arrow] (prompt1) -- (gemini_pro);
        \draw [arrow] (prompt2) -- (gemini_flash1);
        \draw [arrow] (prompt3) -- (gemini_flash2);

        \coordinate (brace_x) at (4.8, 0);

        \draw [decorate, decoration={brace, amplitude=10pt}, thick]
            (corpus.north -| brace_x) -- (repaired_data.south -| brace_x)
            node [midway, right=12pt, font=\bfseries\small] {Stage I};

        \draw [decorate, decoration={brace, amplitude=10pt}, thick]
            (gemini_flash2.north -| brace_x) -- (final.south -| brace_x)
            node [midway, right=12pt, font=\bfseries\small] {Stage II};

    \end{tikzpicture}
}

\captionsetup{justification=raggedright,singlelinecheck=false}

\vspace{0.25cm}
\begin{minipage}{\linewidth}
    \footnotesize
    \justifying
    \begin{spacing}{1}
    \noindent\textit{Notes}: This flowchart represents our LLM-based data pipeline for a given volume. The output is an LLM-generated dataset. After manual data cleaning, we merge all 41 LLM-generated datasets to construct the complete patent dataset including all German patents from 1877 until 1918. Dashed boxes depict our carefully refined prompts, which are shown in \AppendixA. Throughout our pipeline, the temperature parameter is set to 0.0 for all model invocations.
    \end{spacing}
\end{minipage}

\end{figure}
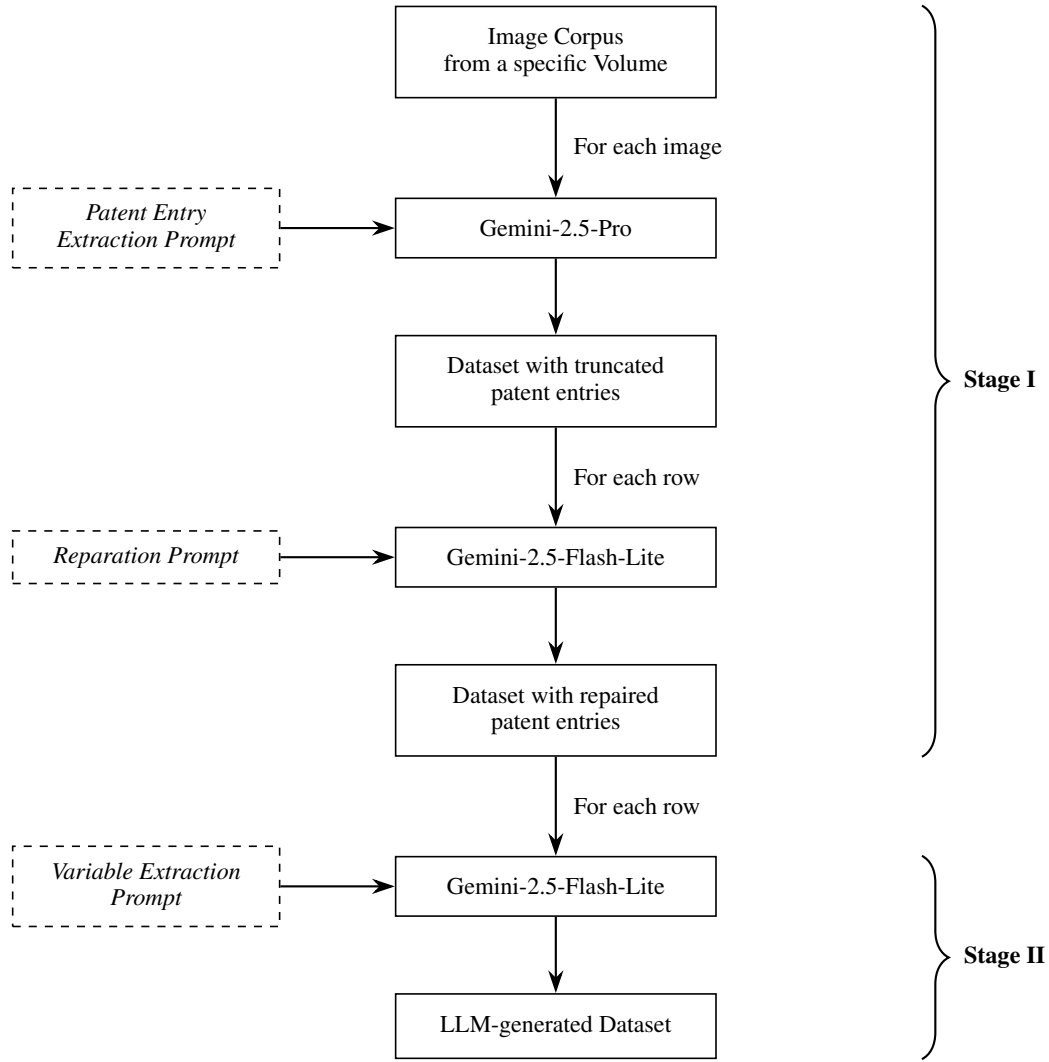

An overview of our LLM-based data pipeline to construct datasets from image scans is depicted in Figure~\ref{fig:pipeline-overview}. In the first stage, we pair each page (e.g., Figure~\ref{fig:representative-page}) with the same prompt and send all prompt-page combinations independently to a multimodal LLM. As output we receive one JSON file per page, containing all patent entries. For a given volume, we concatenate all JSON files and convert them into one large CSV file. At the end of the first stage, our pipeline repairs patent entries that span across pages and columns and returns a dataset in which every observation contains one complete patent entry. In the second stage, we pair each extracted patent entry with a new prompt and send these prompt-entry combinations to another LLM. We receive the extracted variables we specified in our prompt, which are then appended to the corresponding patent entry in our dataset. The following subsections describe the technical details: why we chose the Gemini model family to power our pipeline, how we handled technological classes, repaired patent entries spanning multiple columns or pages, adapted the pipeline to volumes with a special layout, and conducted the manual validation and cleaning of our LLM-generated patent dataset.

\subsection{Model Selection}

We exclusively use models from the Gemini 2.5 family to power our LLM-based data pipeline \citep{comanici2025gemini25pushingfrontier}.\footnote{We do not endorse any mentioned products, services, or organizations, and do not provide legal or financial advice.} Critics may argue that we are using proprietary models offered by a private company. Open-source LLMs do exist, however, initial experiments clearly showed that they are far from capable of extracting the desired information from our image scans. Even if capable open-source alternatives had existed, they would have required access to expensive GPU hardware, which we did not have. Proprietary models from other frontier labs, such as Anthropic or OpenAI, also failed to deliver promising results. Therefore, at the time of coding, Gemini-2.5-Pro was the only model with promising visual capabilities for extracting information from our image corpus. In the end, the majority of economic historians is most likely indifferent about the LLM powering their data pipeline---as long as it quickly produces a high-quality dataset at a cheap price.

\newpage
\newpage

\subsection{Stage I: Patent Entry Extraction from Image Scans}

Our first stage builds upon work by \citet{greif2025multimodal}. Initially, the PDF for a given volume is split into PNGs. Each page is paired with the same prompt (Figure~\ref{fig:01-prompt}) and sent to Gemini-2.5-Pro via the Application Programming Interface (API).\footnote{An API is the standard way to send input to an LLM hosted by a commercial provider in its data centers, which then returns the model's output.} This is done independently for all PNGs at the same time. The prompt is crucial to properly extract the desired information from the archival image scan. We carefully instruct the model to extract all patent entries and technological classes from top to bottom, starting with the column on the left before proceeding with the column on the right-hand side. For each page, we receive a JSON object that contains all patent entries and technological classes in sequential reading order. We retry pages whose API call to the LLM failed until they succeed. Patent entries are always preceded by the key \textit{entry}, whereas technological classes are preceded by \textit{category}. We concatenate all JSON files in document order and keep track of the page number. Consequently, we know that all patent entries appearing between two technological classes must belong to the technological class that precedes them. This allows us to convert the concatenated JSON file into a CSV with three variables: \textit{page}, \textit{entry}, and \textit{category}.

Processing every page independently causes one significant issue. Some patent entries begin at the very bottom of the right-hand column and continue at the top-left on the subsequent page. Similarly, entries occasionally span across both columns on a given image, making it more challenging for Gemini-2.5-Pro to correctly extract the complete text. For this reason, some patent entries are truncated after they were extracted from the image scans. We address this issue by sending each entry to Gemini-2.5-Flash-Lite with a prompt (Figure~\ref{fig:02-prompt}) that checks whether an entry is truncated (1) or not (0). We chose Gemini-2.5-Flash-Lite because it is much cheaper than Gemini-2.5-Pro while being as capable of solving this issue. Finally, we merge truncated entries that span across two pages or columns on the primary source material.

One may ask why we did not just send all pages in one PDF to Gemini-2.5-Pro in order to avoid the issue of entries exceeding a single page. First, the retrieval and reasoning performance of multimodal LLMs declines when more images are provided \citep{wu2025visualhaystacksvisioncentricneedleinahaystack}. Second, the maximum output window of Gemini-2.5-Pro is capped. Therefore, we would have to send all volumes in multiple chunks as they typically consist of hundreds of pages, resulting in the same patent entry continuation problem at the first and last pages of each chunk. Third, by processing each page independently, we do know exactly on which PNG a patent entry is located, facilitating manual data cleaning later on.

\subsection{Stage II: Variable Extraction from Patent Entries}

After the first stage of our LLM-based data pipeline, we now have a variable \textit{entry} in which each row contains a complete patent entry. Conditional on each entry, we extract the variables \textit{patent\_id}, \textit{assignee}, \textit{location}, \textit{title}, and \textit{date} using Gemini-2.5-Flash-Lite. We instruct the model to extract the specified variables, provide some examples, and append the actual patent entry at the bottom of the prompt (Figure~\ref{fig:03-prompt}). In each API call, the model only processes a single, isolated patent entry that was extracted in the first stage. If an API call fails, it is automatically retried up to ten times. Gemini-2.5-Flash-Lite returns a JSON object with keys for the desired variables. If a variable is not present in the patent entry, we instruct the model to return \textit{NaN}. At the end of Stage II, each of the 41 volumes was transformed into a dataset containing the following variables: \textit{page}, \textit{entry}, \textit{category}, \textit{patent\_id}, \textit{assignee}, \textit{location}, \textit{title}, and \textit{date}.

\subsection{Special Volumes with Different Layout}

The first two volumes (1877--8 and 1879) have a slightly different layout than the volumes from 1880 to 1918. The very first volume combines 1877 and 1878 as the Reich Patent Office was established on July 1, 1877. The major difference for these two volumes is that the ID is located at the end of each patent entry and is always preceded by ``P. R.''. Moreover, the combined volume 1877--8 does not contain information on the location of the assignees. Our pipeline allows us to easily adjust the prompts. After experimenting with various prompts on the benchmarking dataset, we amended the prompts across all stages of our pipeline to construct the datasets for the years 1877--8 and 1879. The prompts for these two volumes can be found in \AppendixB.

\subsection{Manual Data Cleaning}

After our LLM-based data pipeline constructed 41 datasets---one for each of the physical volumes---our research assistants manually validated and cleaned a small number of observations. We created auxiliary variables throughout our pipeline to flag truncated entries that were merged, entries for which the API call did not succeed, and samples for which no patent ID was found. Using these auxiliary variables, our research assistants filtered each of our 41 datasets to check, repair, and delete these observations.

\begin{figure}[htpb]

\caption{\\\textsc{INTERFACE FOR ACCELERATED MANUAL DATA CLEANING}}
\label{fig:ra-interface}

\centering
\includegraphics[width=\textwidth]{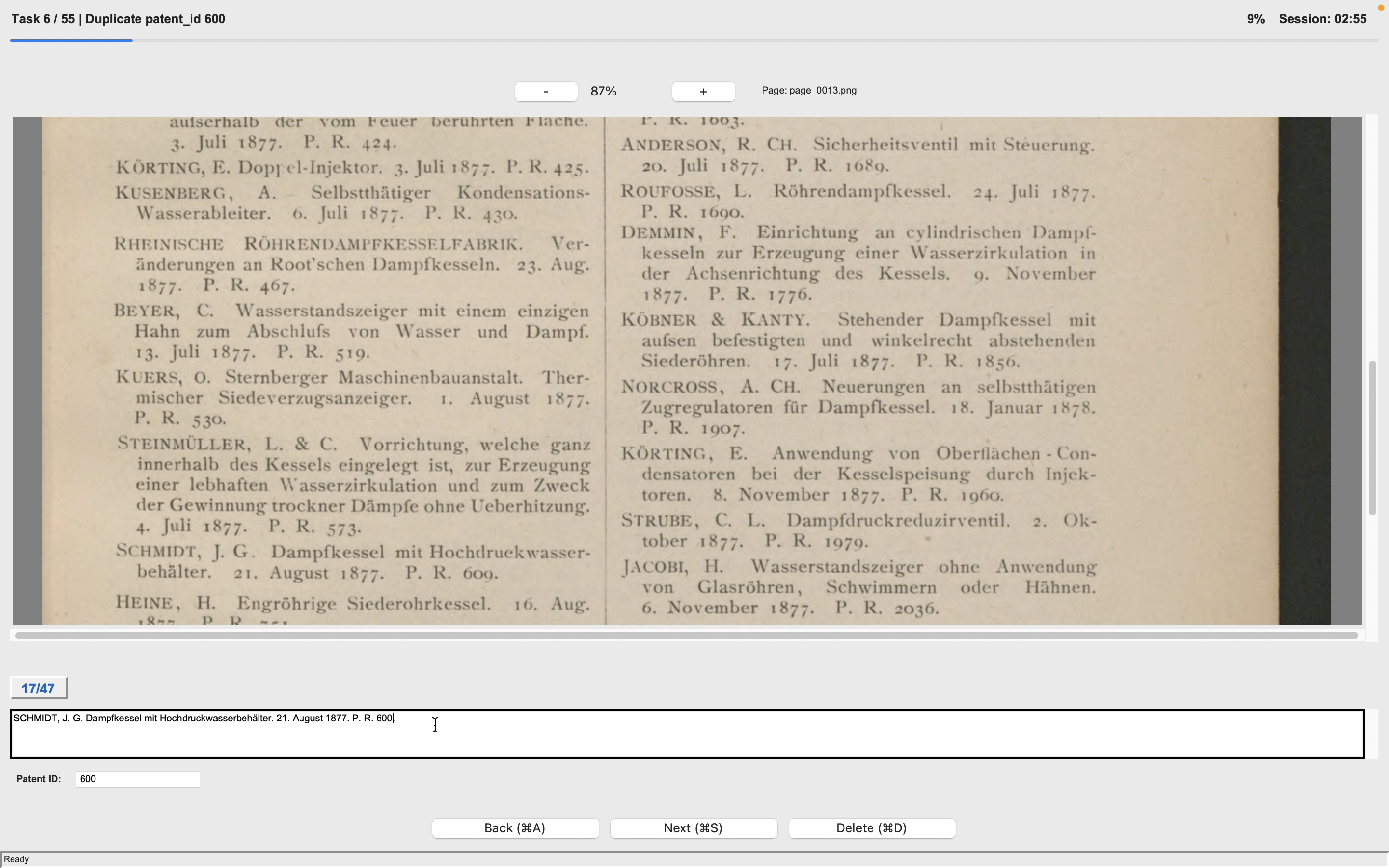}
\captionsetup{justification=raggedright,singlelinecheck=false}

\vspace{0.25cm}
\begin{minipage}{\linewidth}
    \footnotesize
    \justifying
    \begin{spacing}{1}
    \noindent\textit{Notes}: We created this interface using LLM-assisted coding tools. Our research assistants used it to manually clean and validate patent entries with duplicate ID values or IDs that fall outside the volume's range. In the example, the lower part of the ``9'' in patent ID ``609'' is faintly printed, which caused Gemini-2.5-Pro to extract it as ``600''. Because every patent ID is unique, this discrepancy was flagged. Our research assistants then corrected these mistakes by verifying the outputs with their own eyes, whenever applicable. Our interface accelerated this process as our research assistants could make an edit and just had to click ``Next''.
    \end{spacing}
\end{minipage}

\end{figure}

Afterward, we exploited the structure of the primary sources to ensure the quality of each of the 41 datasets. Patents were recorded starting with one-unit increments from patent number $1$. The table of contents indicates the first and last patent number of the corresponding volume, allowing us to check for duplicates and resolve entries with patent numbers that are below or above the volume's range.\footnote{Due to World War I, around 4,000 patent numbers were not entered into the patent register for the volumes 1917 and 1918.} Using LLM-assisted coding tools, we quickly constructed a visual interface for our research assistants (Figure~\ref{fig:ra-interface}). For example, if a patent ID was marked as duplicate, the interface rendered the corresponding page at the top, while the current values for the entry and ID were editable below. Our research assistants then only had to check the image scan and update the patent entry and ID fields, if applicable. Throughout this validation process, we also noticed that the primary source itself contained errors. For example, the patent clerks occasionally made mistakes, resulting in patent entries with duplicate IDs.\footnote{We did not correct these errors made by the employees of the Reich Patent Office and did not assign new, ahistorical patent numbers. Our dataset therefore occasionally contains two different patents with the same patent number.}

The structure of our primary sources also helped us to identify observations in which the technological class was transcribed incorrectly. In a given volume, all patents were listed sequentially with respect to their technological class. For example, class ``2'' must follow class `1''. Similarly, subclass ``17e'' must come after ``17d''. By highlighting inconsistencies among these sequences, our research assistants manually corrected the \textit{category} variable. 

The deviating content of some volumes forced us to auto-delete patent entries in bulk. For the year 1888, we were only able to track down a composite volume that contains all patents that were registered between July 1, 1878, and December 31, 1888, and were still active. Therefore, this volume contains patent entries from preceding volumes, which we removed. Similarly, the volume for year 1879 includes patent entries that were still active and listed in the first volume covering the years 1877 and 1878. These were also removed.

After we conducted the manual data cleaning described above for all 41 LLM-generated datasets, we merged them into a single dataset and prepended the variables \textit{global\_id}, \textit{book}, and \textit{book\_id} to provide high-level metadata.

%% file: 06_benchmarking.tex
\section{Benchmarking}
\label{sec:06_benchmarking}

The benchmarking dataset serves two intertwined purposes. Of course, it allows us to evaluate the quality of our dataset that is produced by our LLM-based data pipeline on a small but representative sample. However, the dataset quality is highly dependent on the prompts we select each time our pipeline employs an LLM. Thus, the benchmarking dataset also enables us to find prompts that maximize the accuracy of the LLM-generated dataset. As we optimize the performance of our LLM-based data pipeline on a representative corpus of 41 randomly selected images, we argue that the evaluation metrics reported on this sample generalize to our full image corpus.

\subsection{Hallucinations}

LLMs have often been caught hallucinating as they can fabricate convincing but incorrect information. In order to assess whether this problem occurs in our task, we had to build a human-made benchmarking dataset with the help of research assistance, which can be used to evaluate the output of multimodal LLMs. 

There are several types of hallucinations. In our dataset construction task, we can only encounter ``input-conflicting hallucinations, where LLMs generate content that deviates from the source input provided by users'' \citep[p.~5]{zhang2025siren}. More precisely, we define a hallucination as any deviation of the \textit{LLM-generated} dataset with respect to the \textit{perfect} benchmarking dataset.

For our research project, not all hallucinations are equally concerning. Falsely recognizing characters or using modern instead of historical letters may be treated as minor hallucinations. These are negligible when it comes to analyzing the historical mass data. More worrisome are moderate hallucinations, such as when the multimodal LLM extracts a technological class header as a patent entry or when the LLM transcribes a full word with modern instead of historical spelling. Major hallucinations are very concerning regarding the dataset quality and its reliability for downstream econometric analysis. This includes completely false or misinterpreted transcriptions of the patent entry and, even worse, invented patent entries that do not appear on the image scan and therefore do not exist.

\subsection{Benchmarking Datasets: Student-Constructed and Perfect}

We randomly selected 41 image scans, one for each calendar year of the annually published patent register, distributed them across two research assistants, and carefully instructed them on how to create the dataset in Excel. In what follows, we refer to the output they initially produced as the \textit{student-constructed} dataset. This process of constructing datasets from archival image scans with the help of research assistants was standard practice for decades in economic history research. Consequently, we are confident that the \textit{student-constructed} dataset approximates the average quality of existing economic history datasets.

After the research assistants produced the \textit{student-constructed} dataset, we generated the corresponding dataset using our LLM-based data pipeline on the 41 sampled images. We then compared the \textit{LLM-generated} dataset with the \textit{student-constructed} dataset to assess the accuracy of our pipeline. We noticed major differences between the two datasets. Initially, we suspected that Gemini-2.5-Pro hallucinated patent entries, but we quickly realized that the research assistants had accidentally skipped some entries. We also observed several transposition errors in which the students extracted ``299187'' as the patent number, but the LLM correctly wrote ``299178''. In addition, we noticed many typos, such as repeated or omitted characters. For this reason, we asked the research assistants to correct their initial \textit{student-constructed} benchmarking dataset until it was \textit{perfect}. To the best of our knowledge, the \textit{perfect} dataset does not contain any remaining errors.

\subsection{LLM-assisted Iterative Prompt Refinement on the Benchmarking Dataset}

Our approach to finding optimal prompts throughout our pipeline resembles contemporaneous work by \citet{backer2025can} and \citet{vafaie2025end}. We iteratively refined the prompts across our whole pipeline with the aim of generating a dataset that matches the \textit{perfect} benchmarking dataset as accurately as possible. After each iteration, we examined our selected evaluation metrics and inspected side-by-side comparisons of the \textit{LLM-generated} and \textit{perfect} datasets. This visual inspection revealed the shortcomings of our current prompts. We described these inaccuracies (e.g., for an early prompt, locations were extracted with the prepositions ``in'' when creating the location variable; ``in Berlin'') and instructed our LLM-assisted coding tool to improve our prompts by taking these issues into account. We carried out several iterations of this LLM-assisted prompt refinement technique. Most likely, there exist other prompts that would yield even higher values on our evaluation metrics. However, after visual inspection, we concluded that our prompts were strong enough to ensure high data quality on our full image corpus. 

We wrote the initial prompts by infusing our historical domain knowledge about the primary source. As a general rule, we started a prompt by assigning a role to the LLM (e.g., ``You are a specialist data extraction AI.''), followed by a brief outline of the general task. We then provided very detailed and explicit instructions, after which we demonstrated some examples with their desired JSON output, if applicable. At the very end, we frequently instructed the models to provide only JSON output and nothing else to avoid unnecessary explanations, which would interfere with the subsequent processing steps in our pipeline, such as concatenating JSON objects or integrating them into the dataset.

\subsection{Character Error Rate}
\label{subsec:cer}

Before benchmarking the Stage I and Stage II of our pipeline, we evaluated the transcription capabilities of Gemini-2.5-Pro by focusing exclusively on the patent entries. Nevertheless, this required performing the first part of Stage I, namely sending the 41 benchmarking images to the model. For each volume, we concatenate all \textit{LLM-generated} patent entries into a single text file to measure the transcription accuracy. We do the same for the \textit{student-constructed} and \textit{perfect} benchmarking datasets.

We then use the character error rate (CER), a standard metric from the field of OCR, to quantify how different two text documents are at the character level. The CER is defined as follows:

\begin{equation}
\label{eq:cer}
\mathrm{CER} = \frac{\text{Levenshtein Distance}}{N} = \frac{S + D + I}{N}
\end{equation}

where the Levenshtein distance is defined as the sum of the number of substitutions $S$, the number of deletions $D$, and the number of insertions $I$. $N$ is the number of characters in the \textit{perfect} text file. Intuitively, the CER is the minimal number of keystrokes needed to turn the \textit{LLM-generated} patent entry transcriptions into the \textit{perfect} text file (or vice versa), normalized by the length of the \textit{perfect} file. 

\begin{figure}[htpb]

\caption{\\\textsc{CHARACTER ERROR RATE BY YEARLY VOLUME: RESEARCH ASSISTANTS VS LLM}}
\label{fig:cer-chart}

\centering
\includegraphics[width=\textwidth]{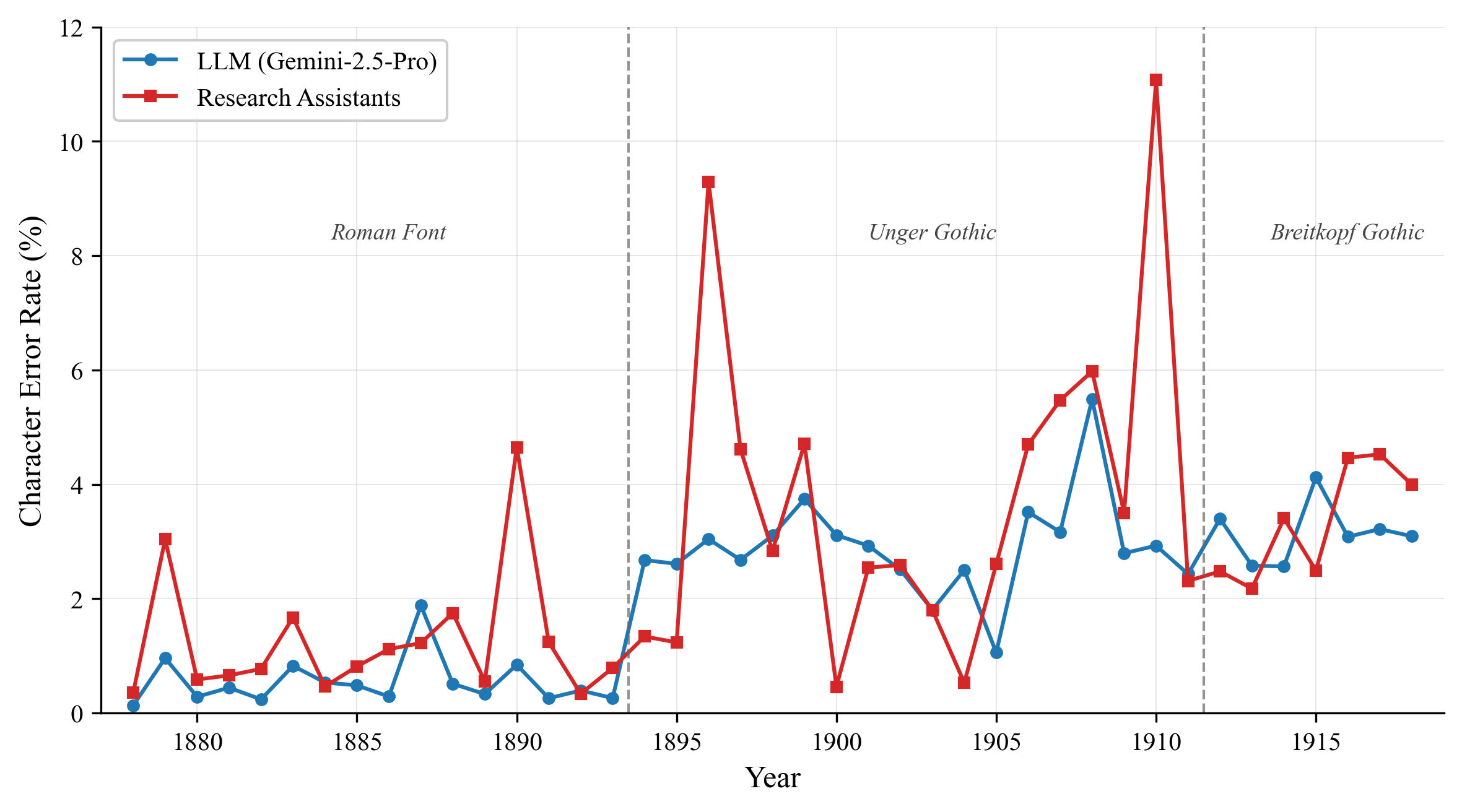}
\captionsetup{justification=raggedright,singlelinecheck=false}

\vspace{0.25cm}
\begin{minipage}{\linewidth}
    \footnotesize
    \justifying
    \begin{spacing}{1}
    \noindent\textit{Notes}: We sampled one page from each of the 41 annual volumes. Each data point is the CER computed over all \textit{LLM-generated} (blue) or \textit{student-constructed} (red) patent entry transcriptions with respect to the \textit{perfect} ones. We did not normalize the transcriptions because we wanted to preserve their historical accuracy. This is an unusual choice and leads to higher CER values, but allows us to see more clearly how close our research assistants and Gemini-2.5-Pro come to the \textit{perfect} transcriptions that are actually depicted on the archival image scans.
    \end{spacing}
\end{minipage}

\end{figure}

Figure~\ref{fig:cer-chart} reports the CER distance both between the \textit{student-constructed} and \textit{perfect} transcriptions and between the \textit{LLM-generated} and \textit{perfect} transcriptions across all 41 volumes. Gemini-2.5-Pro produces transcriptions that are closer to the \textit{perfect} version than the efforts of our research assistants for 27 of the 41 text files. There is one file in which both yield the same CER, while the research assistants are more accurate on the other 13 files. The CER clearly increases for the Gothic fonts, but this appears to hold true for our research assistants as well, who told us that they had to rely on a Gothic reading aid. A closer inspection shows that Gemini-2.5-Pro mainly struggles with the historical long s and occasionally with capitalization.\footnote{We provide an accompanying website that presents all benchmarking results in a visual format, allowing readers to assess the quality of the dataset for themselves. See \href{https://historymind.ai}{\texttt{historymind.ai}}.}

The fact that the CER on our patent entries tends to be lower than that of our research assistants is highly reassuring with regard to severe hallucinations. For example, a single hallucinated patent entry would increase the CER by 2–3\% at the volume level. Nevertheless, we found five notable hallucinations that we wanted to discuss to be fully transparent about the quality of our LLM-generated dataset. The most severe hallucination we detected occurred with our benchmarking image from volume 1878. Gemini-2.5-Pro partially fabricated non-existent parts of the title of the patent with ID 7763. Moreover, the last patent entry on the benchmarking image in volume 1907 was completed by the model with four words that were not depicted in the image. The model also failed to correctly extract the first patent entry in our benchmarking image from volume 1915, as part of the beginning is missing. For the image from volume 1908, the transcription of the patent with ID 199989 is missing an additional priority claim that follows the application date. Finally, on the benchmarking image from volume 1913, the patent entry with ID 259447 reads ``Versteinung'', whereas the LLM incorrectly extracted ``Verkoksung''. All of these examples demonstrate one important point: while Gemini-2.5-Pro tends to perform better than our research assistants at transcribing patent entries, the resulting dataset based on our double-column primary sources remains imperfect.

\subsection{Stage I: Patent Entry Extraction from Image Scans}
\label{subsec:entry-matching}

While the CER provides tentative evidence that Gemini-2.5-Pro can extract and transcribe patent entries more accurately than our human research assistants, we also need to assess how well our LLM-based pipeline can preserve structure and repair patent entries.

We perform one-to-one matching between the datasets containing the \textit{LLM-generated} and \textit{perfect} entries. To allow for small differences caused by misidentified characters, we employ fuzzy string matching. Two patent entries match if they are at least 90\% identical at the character level, which approximately corresponds to allowing errors in no more than one tenth of the characters.\footnote{Greedy one-to-one fuzzy matching was performed using the rapidfuzz library's normalized Levenshtein similarity metric with the threshold set to 0.90.} We perform this matching before and after repairing truncated entries. In this benchmarking exercise, truncated entries can only be repaired when an entry spans columns, because we sampled one page from each volume. This is conceptually identical to truncated entries spanning two subsequent pages, providing a solid testing ground for evaluating the reliability of our reparation procedure.

Table~\ref{tab:patent_entry_matching} depicts the benchmarking results. Directly after the patent entry extraction, 1,360 out of 1,403 \textit{LLM-generated} entries match one of the 1,385 \textit{perfect} entries. 18 of the 41 benchmarking image samples contain a patent entry that spans both columns. Initially, both parts of these spanning entries are extracted as two separate patent entries, which explains why Gemini-2.5-Pro extracted more entries than actually exist. After repairing these spanning entries, the \textit{LLM-generated} benchmarking dataset also consists of 1,385 entries, of which 1,378 have a matching counterpart in the \textit{perfect} benchmarking dataset. Five of the seven unmatched entries are not paired due to the hallucinations described in Section~\ref{subsec:cer}. The remaining two contain a high proportion of the historical long s character but are otherwise largely correct.

\begin{table}[htbp]
    \centering
    \caption{\\\textsc{STAGE I BENCHMARKING RESULTS}}
    \label{tab:patent_entry_matching}
    \vspace{0.2cm}
    
    \begin{tabular}{l ccc}
    \toprule
    & \textbf{Extraction} & \textbf{Reparation} & \textbf{Student-Constructed} \\
    & (Gemini-2.5-Pro) & (Gemini-2.5-Flash-Lite) & \\
    \midrule
    \textit{Perfect} Entries & 1,385 & 1,385 & 1,385 \\
    Extracted Entries & 1,403 & 1,385 & 1,373 \\
    Matched Entries & 1,360 & 1,378 & 1,358 \\
    \% \textit{Perfect} Matched & 98.19\% & 99.49\% & 98.05\% \\
    \% Extracted Matched & 96.94\% & 99.49\% & 98.91\% \\
    \bottomrule
    \end{tabular}
    
    \vspace{0.5em}
    \begin{minipage}{0.82\textwidth}
    \footnotesize
    \textit{Notes:} The \textit{perfect} benchmarking dataset consists of 1,385 patent entries. Column ``Extraction'' refers to the initial part of the first stage where patent entries are extracted from image scans using Gemini-2.5-Pro. Column ``Reparation'' reports performance after repairing truncated entries using Gemini-2.5-Flash-Lite. The third column displays the performance of our research assistants, which is best compared to the ``Reparation'' column. Matches were determined using the rapidfuzz library's normalized Levenshtein similarity metric with the threshold set to 0.90. The reported percentages are also known as Recall (\% \textit{Perfect} Matched) and Precision (\% Extracted Matched).
    \end{minipage}
    \end{table}

\subsection{Stage II: Variable Extraction from Patent Entries}

To benchmark the second stage, we evaluated how accurately Gemini-2.5-Flash-Lite extracts the desired variables from the subsets of patent entries that matched in the previous step (Section~\ref{subsec:entry-matching}). Within those two subsets (\textit{LLM-generated} and \textit{student-constructed}), we extract the desired five variables (\textit{patent\_id}, \textit{assignee}, \textit{location}, \textit{title}, and \textit{date}) for each patent entry and compare them with the corresponding variables in the \textit{perfect} benchmarking dataset.

To determine whether two variable cells match, we apply fuzzy string matching. By choosing the conservative threshold of 90\%, we allow for transcription errors in longer fields, such as patent titles, while still requiring exact correspondence for short fields such as the six-digit patent ID or date.

In Stage I, 1,378 entries match between the \textit{LLM-generated} and \textit{perfect} benchmarking dataset, yielding 6,890 variable cells, of which 6,550 match in Stage II of our benchmarking exercise, a percentage of 95.07\% (Table~\ref{tab:variable_extraction}). The shorter patent ID and date field have an accuracy of 99.49\% and 98.91\%, respectively. The accuracy is lower for assignees (92.16\%), locations (91.36\%), and patent titles (93.40\%), which is expected due to transcription errors from the long s and capitalization that patent IDs and dates do not usually have.

Notably, the \textit{student-constructed} benchmarking dataset only had 1,358 matching entries with respect to the \textit{perfect} dataset because the students completely overlooked some patent entries or incorrectly repeated patent titles from previous entries, among other things. Therefore, the \textit{student-constructed} dataset is evaluated on a smaller subset, making a direct comparison with the \textit{LLM-generated} dataset more difficult. Nevertheless, even on this reduced subset, the research assistants were less accurate in extracting patent IDs, locations, and dates. They were better in extracting assignees and titles. This pattern is explained by the fact that we explicitly instructed our research assistant to use the historical long s, which Gemini-2.5-Pro failed to reproduce.

Some economic historians may argue that the reported variable-level accuracies on the \textit{LLM-generated} dataset are insufficient, or that the remaining errors are systematically biased. For this reason, we closely examined the non-matching variable cells.\footnote{All variable extraction benchmarking results are visually available on our paper website at \href{https://historymind.ai}{\texttt{historymind.ai}}.} For the assignee field, Gemini-2.5-Flash-Lite sometimes incorrectly included the occupation, which occasionally followed an individual name. Furthermore, Gemini-2.5-Flash-Lite has consistently separated each of the names of three or more patent holders with ``und'' (German for ``and''), while the perfect benchmarking dataset follows a comma-separated listing convention, with ``und'' only appearing before the last name. The \textit{LLM-generated} locations often include a subsequent end point without spaces, which is not present in the \textit{perfect} benchmarking dataset (e.g., \textit{LLM-generated }``Berlin.'' and \textit{perfect} ``Berlin'' do not match).\footnote{Further refining our prompts may have resolved these issues.} Moreover, many locations contain the historical long s. These single character deviations can cause a non-match as many of the extracted locations are short (e.g., ``Dresden'' does not match its counterpart with the historical long s). Based on our benchmarking results, we conclude that our \textit{LLM-generated} patent dataset is of sufficiently high quality for use in economic history research.

\begin{table}[htbp]
    \centering
    \caption{\\\textsc{STAGE II BENCHMARKING RESULTS}}
    \label{tab:variable_extraction}
    \vspace{0.2cm}
    
    \begin{tabular}{l ccc ccc}
    \toprule
    & \multicolumn{3}{c}{\textbf{Gemini-2.5-Flash-Lite}} & \multicolumn{3}{c}{\textbf{Student-constructed}} \\
    \cmidrule(lr){2-4} \cmidrule(lr){5-7}
    & Total & Matched & Match Rate & Total & Matched & Match Rate \\
    \midrule
    Total Cells & 6,890 & 6,550 & 95.07\% & 6,790 & 6,503 & 95.77\% \\
    \midrule
    \multicolumn{7}{l}{\textit{By Variable}} \\
    \midrule
    \quad Patent ID & 1,378 & 1,371 & 99.49\% & 1,358 & 1,329 & 97.86\% \\
    \quad Assignee & 1,378 & 1,270 & 92.16\% & 1,358 & 1,289 & 94.92\% \\
    \quad Location & 1,378 & 1,259 & 91.36\% & 1,358 & 1,231 & 90.65\% \\
    \quad Title & 1,378 & 1,287 & 93.40\% & 1,358 & 1,316 & 96.91\% \\
    \quad Date & 1,378 & 1,363 & 98.91\% & 1,358 & 1,338 & 98.53\% \\
    \bottomrule
    \end{tabular}
    
    \vspace{0.5em}
    \begin{minipage}{0.72\textwidth}
    \footnotesize
    \textit{Notes:} This table reports variable extraction performance comparing \textit{LLM-generated} and \textit{student-constructed} variables against the variables from the \textit{perfect} benchmarking dataset. Based on the benchmarking results in Table \ref{tab:patent_entry_matching}, we retain 1,378 \textit{LLM-generated} and 1,358 \textit{student-constructed} entries, resulting in 6,890 and 6,790 variable cells. For each patent entry, we extract five variables (\textit{patent\_id}, \textit{assignee}, \textit{location}, \textit{title}, and \textit{date}) and compare them with the variable values in the \textit{perfect} benchmarking datasets. Matches between two cells were determined using the rapidfuzz library's normalized Levenshtein similarity metric with the threshold set to 0.90.
    \end{minipage}
    \end{table}

%% file: 07_economics_of_llms.tex
\section{The Economics of Constructing Datasets from Image Scans Using Multimodal LLMs}
\label{sec:07_economics_of_llms}

Until now, economic historians around the world have been constructing datasets from primary sources by hand. The task of extracting information from archival sources is typically conducted by research assistants, PhD students, and sometimes even outsourced to specialized companies. This process is very time-consuming and costly. When we constructed our benchmarking datasets, we asked our research assistants to track their time. On average, they worked roughly two hours per page to produce the \textit{student-constructed} dataset, including the full patent entries as well as the five desired variables. Creating the \textit{perfect} benchmarking dataset required an additional two hours per page. As our image corpus corpus consists of 9,562 pages, it would have taken about 19,124 hours of human labor to construct the patent dataset using the standard practices in economic history. By contrary, our LLM-based data pipeline can generate the full dataset within a day. If we assume a German minimum wage of EUR 12.82, the construction of the patent dataset would have cost at least EUR 245,269. By contrast, the LLM-based data pipeline cost EUR 1,196 (Table~\ref{tab:cost_breakdown}). Consequently, multimodal LLMs enabled us to construct the patent dataset more than 795 times faster and 205 times cheaper compared to the standard practice of constructing datasets in economic history.\footnote{Research assistants hired in Baden-Württemberg, Germany currently earn approximately one euro above the minimum wage. Moreover, our pipeline could process all 9,562 image scans in a few seconds. However, we required roughly a single day because of the constraints of our personal computer and to avoid rate limit issues with the Google Gemini API. Consequently, our comparative calculations should be viewed as lower-bound estimates.} The superiority of multimodal LLMs for historical data construction becomes even more pronounced by the fact that the pipeline yielded higher data quality than we would expect from our research assistants.

The primary cost driver in our LLM-based data pipeline is the model selection. To ensure the highest data quality, we could have exclusively used the most capable model of the Gemini family, Gemini-2.5-Pro at the time of coding, across all points where LLMs are employed in our pipeline. However, this would have led to much higher costs with only slight quality improvements. For this reason, we chose the less capable but cheaper Gemini-2.5-Flash-Lite for repairing truncated entries and extracting variables from the previously extracted patent entries. The model exhibited satisfactory performance on the benchmarking dataset, so why should we use a more intelligent and expensive model if the results are already satisfactory? As the cost of LLM inference continues to fall quickly \citep{gundlach2025priceprogressalgorithmicefficiency}, researchers have to constantly reevaluate which models meet their requirements, depending on their digitization project. Once open-source models reach the performance of current frontier models, they may become reliable alternatives for dataset construction tasks.

\begin{table}[htbp]
\centering
\caption{Cost Breakdown by Pipeline Stage}
\label{tab:cost_breakdown}

\vspace{0.2cm}

\begin{tabular}{l l rr rr r r}
\toprule
& & \multicolumn{2}{c}{\textbf{\$/1M Tokens}} & \multicolumn{2}{c}{\textbf{Tokens Used}} & & \\
\cmidrule(lr){3-4} \cmidrule(lr){5-6}
\textbf{Stage} & \textbf{Model} & Input & Output & Input & Output & \textbf{Cost (\$)} & \textbf{\%} \\
\midrule
\multicolumn{8}{l}{\textit{Stage I: Patent Entry Extraction from Image Scans}} \\
\midrule
\quad Extraction & Gemini-2.5-Pro & \$1.25 & \$10.00 & 24.1M & 107.6M & {1,105.9} & 92.4\% \\
\quad Reparation & Gemini-2.5-Flash-Lite & \$0.10 & \$0.40 & 183.1M & 0.3M & 18.4 & 1.6\% \\
\midrule
\multicolumn{8}{l}{\textit{Stage II: Variable Extraction from Patent Entries}} \\
\midrule
\quad Extraction & Gemini-2.5-Flash-Lite & \$0.10 & \$0.40 & 612.9M & 26.6M & 71.9 & 6.0\% \\
\midrule
\textbf{Total} & & & & \textbf{820.1M} & \textbf{134.5M} & \textbf{1,196.3} & \textbf{100\%} \\
\bottomrule
\end{tabular}

\vspace{0.5em}
\begin{minipage}{0.87\textwidth}
\footnotesize
\textit{Notes:} We tracked all costs within our LLM-based data pipeline using prices as of August 2025. We did not use batch processing, which would have reduced the overall cost by 50\%. Output tokens include both thinking (reasoning) and output tokens. The major cost driver of our pipeline are the 107.6 million output tokens of Gemini-2.5-Pro. All token counts are rounded to one decimal place.
\end{minipage}
\end{table}

The cost to use an LLM is determined by the length of the input it receives and by the length of the output it produces. The length is measured in tokens. A text token can be a single word or a smaller unit of text. Analogously, images are converted to image tokens. In a single request to Gemini-2.5-Pro, the image scan and the prompt are tokenized and jointly represent the total input tokens. In general, input tokens are cheap, while output tokens are much more expensive. Due to these underlying economics, writing a very explicit prompt does not incur much cost. In contrast, Gemini-2.5-Pro's output tokens are very expensive, explaining why 92.4\% of our budget was spent on extracting the patent entries from our image scans. Nevertheless, the patent entry extraction from the archival image scans is the foundation of the whole dataset quality, which is the reason why we wanted to use the best multimodal LLM available at the time of implementation. In addition to input and output tokens, Gemini-2.5-Pro and Gemini-2.5-Flash-Lite also generate thinking tokens. Upon a request, both models first ``think and reason'' before providing an answer. With increasing task difficulty, the models spend more resources on thinking tokens in order to outline and plan how to successfully accomplish the user request. These thinking tokens are charged at the same price as output tokens, which further adds to the high cost of our first stage.

%% file: 08_conclusion.tex
\section{Conclusion}
\label{sec:08_conclusion}

In this paper, we have made several contributions. First, we built and published an LLM-based data pipeline that made it possible to transform the image scans of the German historical patent register into a structured patent database that provides information about the patent holders and the content of all 306,070 patents of the German Empire. Second, we created and open-sourced a \textit{student-constructed} and a \textit{perfect} benchmarking dataset to evaluate the accuracy of multimodal LLMs on the task of historical dataset construction from image scans. Our benchmarking exercise provides tentative evidence that multimodal LLMs can generate datasets of higher quality than those constructed by human research assistants. While we can only be certain about this statement on our image corpus, there is strong reason to believe that this may also be the case for other and even more complex historical sources. Therefore, we encourage other economic historians to use LLM-assisted coding tools to adopt our pipeline to their image corpora in order to explore the general reliability of multimodal LLMs. Third, we open-sourced the full patent dataset, an invaluable source to study the history of innovation. In the remainder of this conclusion, we will discuss the potential impact of multimodal large language models on the field of economic history.

For a long time, economic history research suffered from a lack of large-scale microeconomic data. The advent of multimodal large language models might put an end to this problem. However, a systematic risk in deploying AI agents to construct large-scale datasets is the infeasibility of validating all data entries, which is exacerbated by our limited understanding of LLMs. The potential introduction of unnoticed hallucinations---or even worse---of systematic bias into LLM-generated datasets threatens the robustness of the subsequent econometric analysis. This can have harmful real-world consequences if insights based on biased datasets shape modern policy. Our LLM-generated dataset, for example, has 306,070 patent entries, each yielding the five extracted variables and the technological class, making a manual review of all 2,142,490 data cells practically impossible. Therefore, we have to rely on the results reported on our benchmarking dataset and assume that the performance on this representative subset of 41 pages will generalize to our full image corpus. The reliability of future patent-based historical research on the causes and consequences of German innovations is based on the accuracy of this assumption. Finally, we note that this new way of generating mass data departs from the standard practice in economic history, where close manual work has has helped economic historians to acquire a deep historical understanding and intuition regarding the primary source material.

Multimodal LLMs may soon allow economic historians to build datasets from image scans on demand by simply sending a PDF to a model with the instruction to output the dataset as CSV or JSON object.\footnote{During writing, Gemini-3-Pro-Preview was released in November 2025. This model shows even stronger visual abilities, and when prompted with ``Create an economic history dataset'' and given a multi-page PDF, it seems to output a near-perfect dataset while being able to handle truncated patent entries. Gemini-3-Flash-Preview was released shortly after in December 2025 and outperforms Gemini-2.5-Pro across almost all benchmarks while being more than three times cheaper. We initially developed the two-stage pipeline because, with Gemini-2.5-Pro, the best strategy was to keep dataset construction tightly controlled by breaking the task into smaller subproblems rather than asking the model to generate full datasets directly from one or multiple image scans. Therefore, we reserve the right to rebuild the dataset in the future with an even stronger model to increase dataset quality further and to rule out any remaining hallucinations.} Multimodal LLMs will also enable the research community to quickly replicate well-known economic history datasets and test whether the downstream analyses are robust. On-demand dataset construction from image scans may also steer the incentives of scholars toward open-sourcing the datasets they painstakingly created by hand over decades. These datasets were often kept private, as economic historians planned to use them for their own publications. However, on-demand dataset construction may also incentivize public digital archives to take down their collections of archival image scans, aiming to prevent others from downloading them and creating datasets on which they could then publish their own research.

Future research should benchmark how LLMs perform when constructing  historical datasets based on handwritten fonts and low-resource languages, or when provided  with a PDF containing multiple pages instead of processing each page independently. Furthermore, attempts should be made to implement text-as-data as image-as-data approaches. Why should economic historians work with editable text files when they can work directly with archival image scans depicting embedded text? In general, economic historians should strive to work with the closest representation of the primary source. For example, consider a paper that aims to determine the sentiment transmitted in every daily BBC television broadcast since 1936. Instead of relying on audio transcripts, economic historians should use multimodal LLMs that can directly process videos, which may capture more nuances in how information was conveyed to the nation. Economic history, like other social sciences and humanities, may enter an age of data abundance, in which economic historians can answer new questions simply by constructing new large-scale datasets from sources such as our patent statistics.

Multimodal LLMs may not change what is at the core of economic history, but they will accelerate the research cycle, as has already been the case in other domains \citep{bubeck2025earlyscienceaccelerationexperiments}. For example, identifying gaps in the literature and then proposing promising hypotheses has already yielded successful results in the biomedical domain \citep{gottweis2025aicoscientist}. Moreover, there are ongoing efforts to build AI Scientists that automate data-driven scientific discovery \citep{mitchener2025kosmosaiscientistautonomous}. In economics, there is an emerging literature on AI agents to accelerate economic research \citep{NBERw34202}. However, such debates are almost absent in economic history, even though the field is even more susceptible to automated data-driven discovery, as all the answers are, in principle, based on the surviving documents and artifacts. While many questions about the future development and impact of AI remain uncertain, we are convinced that the nature of work in economic history will change. The tasks of research assistants will shift away from manually constructing historical datasets from primary sources to orchestrating LLM-based data pipelines with LLM-assisted coding tools and managing the underlying data infrastructure. Economic historians are no longer limited by manual dataset construction and will have to learn to navigate this abundance of data to filter out those research questions that promise new and important insights.

%% file: 09_appendix_a.tex
\section*{Appendix A: Prompts}
\label{sec:appendix_a}
\setappendixfigures{A}

\begin{center}
    \captionsetup{hypcap=false}
    \captionof{figure}{\\\textsc{Patent Entry Extraction Prompt}}
    \label{fig:01-prompt}
\end{center}

\vspace{-1.5em}

\begin{tcolorbox}[
    breakable,
    enhanced,
    colback=gray!5!white,
    colframe=black,
    coltitle=white,
    fonttitle=\bfseries,
    boxrule=0.5pt,
    arc=0mm,
    top=0mm, bottom=0mm, left=0mm, right=0mm,
    boxsep=1mm
]
\begin{lstlisting}[
    basicstyle=\scriptsize\ttfamily,
    breaklines=true,
    breakatwhitespace=true,
    columns=fullflexible,
    keepspaces=true,
    aboveskip=5pt, belowskip=5pt,
    % --- THE FIX: EXTENDED CHARACTER MAPPING ---
    literate=
        {ö}{{\"o}}1
        {ä}{{\"a}}1
        {ü}{{\"u}}1
        {Ö}{{\"O}}1
        {Ä}{{\"A}}1
        {Ü}{{\"U}}1
        {ß}{{\ss}}1
        {—}{{---}}1   % Em-dash (long)
        {–}{{--}}1    % En-dash (medium) - This was likely causing the "Vgl." crash
        {−}{{--}}1    % Minus sign (math)
        {†}{{\dag}}1  % Dagger
        {ſ}{{s}}1     % Long s (mapped to normal s)
        {А}{{A}}1     % Cyrillic A (looks like Latin A, common in OCR) -> Latin A
]
You are a specialist data extraction AI. Your task is to process a single page from a German Imperial Patent Office register and extract the specified contents into a perfectly structured, flat JSON array. You must act with the precision of a world-class quantitative historian, ensuring absolute accuracy and adherence to the following rules. **HISTORICAL ACCURACY IS PARAMOUNT - extract everything exactly as written, preserving all abbreviations, spellings, and punctuation exactly as they appear in the original text.**

## I. Core Principles & Scope

### Extraction Area
You will **ONLY** extract information from the two main content columns on the page.

### IGNORE Absolutely:
*   **Running Page Headers:** The text at the very top of the page (e.g., "Kl. 18. Nr. 55711." or "Kl. 19. Nr. 57 185." at the top-left and top-right).
*   **Page Numbers:** Any numbers located in the center of the running header.
*   **Footer Information:** Any text at the bottom of the page that begins with an asterisk (*), is a correction ("Berichtigung"), or contains standalone numbers (e.g., "2*", "1917") that are clearly not part of the patent entries above.

### Reading & Processing Order
Your output must strictly follow this sequence:
1.  First, process the **LEFT** content column from top to bottom.
2.  Second, process the **RIGHT** content column from top to bottom.

The final JSON array must reflect this sequential order of extracted items.

## II. Extraction Rules: Identifying & Formatting Content

For each block of text you encounter in the two main content columns, you must classify it as one of the following and format it accordingly.

### A. Category Headings

*   **Identification:** A category heading is a line that signals a new patent class or subclass.
    *   A main class heading always starts with "Klasse", followed by a number and a title (e.g., "Klasse 19. Eisenbahn- und Brückenbau.").
    *   A subclass heading is typically just the class code itself, often on its own line (e.g., "15a." or "45i.").
*   **Extraction Rule:** From the identified heading, you must extract **ONLY** the class number/code and nothing else.
    *   If the heading is "Klasse 19. Eisenbahn- und Brückenbau", you will extract "19".
    *   If the heading is "18 b.", you will extract "18b".
    *   Clean the extracted code by removing any trailing periods or whitespace.
*   **JSON Output:** The extracted class code must be placed in an object with the key "category".
    > **Example:** {"category": "3"}
    > **Example:** {"category": "3a"}

### B. Patent Entries

*   **Identification of a Standard Entry:** A standard, complete patent entry is a paragraph that:
    *   Begins with a patent ID number (e.g., 55711.).
    *   The ID may be prefixed with an enumeration (e.g., 1.) or a dagger symbol in parentheses (e.g., (†)).
    *   Contains descriptive text about the patent holder, location, title, description, etc.
    *   Typically ends with a date and a code (e.g., 26. Februar 1890. A — 252.).
    *   **Very rarely, additional information may appear after the date/code (e.g., "Priorität aus der Anmeldung in Österreich vom 23/5 08 anerkannt.") - include this if present.**

*   **Extraction Rule for ALL Entries (Standard and Truncated):**
    *   Capture the text of the entry verbatim, preserving all original characters, spelling, and punctuation.
    *   **Line Break Handling:**
        *   Join words that are hyphenated across a newline (e.g., Hindersin-\nstraſse becomes Hindersinstraſse).
        *   Replace all other newlines within the entry's paragraph with a single space to form one continuous line of text.

*   **CRITICAL EXCEPTION: Truncated Entries**
    The single-page context means you will encounter incomplete entries. You **MUST** extract these as they appear.
    *   **Entry Truncated at the Top-Left:** If a column begins mid-paragraph (continuing from a previous page), extract that partial entry as-is.
    *   **Entry Truncated at the Bottom-Right:** If an entry is cut off at the bottom of a column (continuing to the next page), extract that partial entry as-is.
    *   **Entry Spanning Columns on the SAME Page:** If an entry starts at the bottom of the left column and its text continues at the top of the right column, you must create **TWO SEPARATE** entries in the JSON array. The first object will contain the part from the left column, and the second object will contain the part from the right column.

*   **JSON Output:** The formatted text of the entry must be placed in an object with the key "entry".
    > **Example:** {"entry": "55711. COOMES, M, F., Arzt, und A. W. HYDE, in Louisville, 6281/2 Fourth Street bezw. 828 Second Street, Grafschaft Jefferson, Staat Kentucky, V. St. A.; Vertreter: C. PIEPER in Berlin N.W., Hindersinstraſse 3. Verfahren zur Bereitung von Stahl. 26. Februar 1890. A — 252."}

### C. Content to Explicitly IGNORE
*   **Lists of Patent Numbers:** After a category heading, you will often see a block of text that is just a list of patent numbers and codes (e.g., 54571 A 3. — 54814 A 77. — 54881 A 37. — 54977 A 54.). **DO NOT** extract these lists that are just consisting of patent numbers and patent codes. These invalid patent entries only occur directly below category headings. **These lists can span multiple lines and form large blocks of text - ignore the entire block.** Dates in a new line below a valid patent entry belong to the valid patent entry.
*   **Reference Information ("Vgl."):** Any text that begins with "Vgl." (e.g., "Vgl. Kl. 12 P. R. 21188. – Kl. 45 P. R. 17470, 20344.") that appears between patent entries and class headings. **DO NOT** extract this cross-reference information.

## III. JSON Output Specification (Strict)

Your entire output must be a single, syntactically perfect JSON array. Any deviation will result in failure.

*   **Array Structure:** The output must begin with [ and end with ].
*   **Object Structure:** Each element in the array is an object {}.
*   **Key-Value Pairs:** Each object must contain exactly **one** key-value pair. The key must be either "category" or "entry".
*   **Quotation:** **ALL** keys **MUST** be enclosed in double quotes (e.g., "entry"). All string values must be enclosed in double quotes.
*   **Delimiters:** Objects must be separated by a comma ,. The last object in the array must **NOT** have a trailing comma.
*   **Escaping:** Within a string value, any literal double quote " must be escaped as \".

---

**Example of Perfect Output Structure:**

[
  {
    "category": "18"
  },
  {
    "entry": "55711. COOMES, M, F., Arzt, und A. W. HYDE in Louisville, 6281/2 Fourth Street bezw. 828 Second Street, Grafschaft Jefferson, Staat Kentucky, V. St. A.; Vertreter: C. PIEPER in Berlin N.W., Hindersinstraſse 3. Verfahren zur Bereitung von Stahl. 26. Februar 1890. A — 252."
  },
  {
    "entry": "56181. VERSEN, B., in Dortmund, Heiligerweg 8. Verfahren und Vorrichtung zur Herstellung von Bessemer-Birnen-Böden. (Zusatz zum Patente Nr. 30634.) 15. Mai 1890. A — 363."
  },
  {
    "entry": "56195. ADAMS, CH., in Pittsburgh, 110 Diamond Street, Pennsylvania, V. St. A.; Vertreter: F. EDMUND THODE & KNOOP in Dresden, Amalienstraſse 5. Unmittelbare Darstellung von Eisen aus seinen Erzen. 9. Juli 1890. A — 461."
  },
  {
    "entry": "56205. — HERBERTZ, F. A., in Köln a. Rh. Schmelzofen mit Dampfstrahl. 12. August 1890. А — 461."
  },
  {
    "category": "3a"
  },
  {
    "entry": "(†) 55539. — FÜRSTENHEIM, C., in Berlin C., Jerusalemerstraſse 15. Zusammenklappbares Büstengestell. 5. Juni 1890. A — 249."
  }
]

**Final Instruction:**  Adhere to all rules. Your output must be **ONLY the JSON array** and nothing else. Before finishing, double-check your output to ensure it is perfectly valid JSON according to the specifications above.
\end{lstlisting}
\end{tcolorbox}

\begin{minipage}{\linewidth}
    \footnotesize
    \justifying
    \begin{spacing}{1}
    \noindent\textit{Notes}: The prompt was sent together with an image of our image corpus to Gemini-2.5-Pro in order to extract the patent entries and technological classes. Gemini-2.5-Pro returned a JSON object with keys \textit{entry} and \textit{category} in sequential reading order.
    \end{spacing}
\end{minipage}

\begin{center}
    \captionsetup{hypcap=false}
    \captionof{figure}{\\\textsc{Reparation Prompt}}
    \label{fig:02-prompt}
\end{center}

\vspace{-1.5em}

\begin{tcolorbox}[
    breakable,
    enhanced,
    colback=gray!5!white,
    colframe=black,
    coltitle=white,
    fonttitle=\bfseries,
    boxrule=0.5pt,
    arc=0mm,
    top=0mm, bottom=0mm, left=0mm, right=0mm,
    boxsep=1mm
]
\begin{lstlisting}[
    basicstyle=\scriptsize\ttfamily,
    breaklines=true,
    breakatwhitespace=true,
    columns=fullflexible,
    keepspaces=true,
    aboveskip=5pt, belowskip=5pt,
    % --- CHARACTER MAPPING ---
    literate=
        {ö}{{\"o}}1
        {ä}{{\"a}}1
        {ü}{{\"u}}1
        {Ö}{{\"O}}1
        {Ä}{{\"A}}1
        {Ü}{{\"U}}1
        {ß}{{\ss}}1
        {—}{{---}}1   % Em-dash
        {–}{{--}}1    % En-dash
        {−}{{--}}1    % Minus sign
        {†}{{\dag}}1  % Dagger
        {ſ}{{s}}1     % Long s
        {„}{{,,}}1    % German opening quote
        {“}{{''}}1    % German closing quote
]
You are a data validator. Your task is to classify a German patent entry as either valid (`1`) or truncated (`0`).

**CRITICAL DISTINCTION:**
- A **valid** entry contains the END of the patent with date/registration information, regardless of whether it starts from the beginning or middle of the original patent text
- A **truncated** entry starts normally with a patent ID but is cut off mid-sentence and MISSES the end/date

**EXAMPLES:**

**TRUNCATED (tag as `0`)** - These start normally but are cut off mid-sentence:
- "(†) 15322. — SCHWINTZER & GRÄFF in Berlin S., Sebastianstr. 18. Neuerungen an Hänge- und Steh-Schiebelampen, bestehend in einem gläsernen, mit leicht zerlegbarer Metallumkleidung"
- "240938. Sulzer, Robert, Winterthur, Schweiz; Vertr.:"

**VALID (tag as `1`)** - These contain the end with date/registration, even if they start mid-sentence:
- "Neuerungen in der Herstellung der sog. Rahmen für Schuhe und Stiefel. 28. Januar 1882. — A 648."
- "der Ränder des Laufmantels von Luftradreifen. 17/8 13. — A 1492."
- "zigerstr. 91. Zweibehälter-Drahtziehmaschine. 29/1 96."
- "mit auswechselbaren, ineinandergreifenden Segmenten mit seitlichen Hohlräumen und einem mittleren, über den seitlichen Hohlräumen angeordneten, unterstützten Hilfshohlraum. 2/3 13. — A 2069. Priorität aus der Anmeldung in den Vereinigten Staaten von Amerika vom 27/7 12 anerkannt."

Respond with a single character only:
- `1` if the entry contains the end with date/registration information
- `0` if the entry is truncated mid-sentence and missing some part at the end

Do not provide any explanation or additional text.
\end{lstlisting}
\end{tcolorbox}

\begin{minipage}{\linewidth}
    \footnotesize
    \justifying
    \begin{spacing}{1}
    \noindent\textit{Notes}: This prompt was used to identify entries that were truncated after Gemini-2.5-Pro extracted the patent entries from the image scans. An extracted patent entry was appended to the bottom of this prompt and then sent to Gemini-2.5-Flash-Lite. This was done independently for all patent entries. The models returns ``valid'' (1) for any entry containing the closing date and registration code, even if the beginning is missing, while the model returns ``truncated'' (0) for entries that start properly but are cut off mid-sentence. Afterward, our pipeline merged truncated entries with the respective entry below.
    \end{spacing}
\end{minipage}

\begin{center}
    \captionsetup{hypcap=false}
    \captionof{figure}{\\\textsc{Variable Extraction Prompt}}
    \label{fig:03-prompt}
\end{center}

\vspace{-1.5em}

\begin{tcolorbox}[
    breakable,
    enhanced,
    colback=gray!5!white,
    colframe=black,
    coltitle=white,
    fonttitle=\bfseries,
    boxrule=0.5pt,
    arc=0mm,
    top=0mm, bottom=0mm, left=0mm, right=0mm,
    boxsep=1mm
]
\begin{lstlisting}[
    basicstyle=\scriptsize\ttfamily,
    breaklines=true,
    breakatwhitespace=true,
    columns=fullflexible,
    keepspaces=true,
    aboveskip=5pt, belowskip=5pt,
    % --- CHARACTER MAPPING ---
    % Includes the previous chars + German quotes found in your new text
    literate=
        {ö}{{\"o}}1
        {ä}{{\"a}}1
        {ü}{{\"u}}1
        {Ö}{{\"O}}1
        {Ä}{{\"A}}1
        {Ü}{{\"U}}1
        {ß}{{\ss}}1
        {—}{{---}}1   % Em-dash
        {–}{{--}}1    % En-dash
        {−}{{--}}1    % Minus sign
        {†}{{\dag}}1  % Dagger
        {ſ}{{s}}1     % Long s
        {„}{{,,}}1    % German opening quote
        {“}{{''}}1    % German closing quote
]
You will be provided with a single text entry from Germany's Imperial Patent Office. Your task is to function as a highly precise information extraction engine. You must carefully analyze the entry and extract specific fields, formatting the output as a single, valid JSON object.

**Core Principles:**
*   **Exactness:** All extracted text must be an *exact, character-for-character copy* from the source text. Do not add, remove, or alter any characters, including historical German characters (e.g., ſ).
*   **Structure:** The patent entries almost always follow a specific order: `patent_id` -> `name(s)` -> `location(s)` -> (optional `Vertreter` info) -> `description` -> date -> (optional addendum note or priority registration). Use this predictable structure to guide your extraction.
*   **Output Format:** The output MUST be a single JSON object with the keys: `patent_id`, `name`, `location`, `description`, and `date`. If a value cannot be reliably found for a field, use the string "NaN".

---

**Detailed Extraction Rules:**

**1. `patent_id` (String)**
*   **Position:** The patent ID is the numerical identifier at the beginning of the entry.
*   **Format:** It is a one- to six-digit number.
*   **Rule:** You MUST ignore any non-digit prefixes. This includes list enumerations (e.g., `1.`, `17.`), symbols (e.g., `(†)`), dashes, or spaces. Extract ONLY the numeric digits.
*   **Example:** For `(†) 2. 35321. — ...`, you must extract `"35321"`.

**2. `name` (String)**
*   **Position:** This is the name of the patent holder(s), which immediately follows the `patent_id`.
*   **Rule:** Extract the full name(s) exactly as written, but follow these CRITICAL rules:
    *   **INCLUDE:** Corporate forms (e.g., `AKTIEN-GESELLSCHAFT FÜR BERGBAU- UND ZINKHÜTTENBETRIEB`).
    *   **INCLUDE:** All parts of compound names (e.g., `SCHIFFER & KIRCHER`).
    *   **INCLUDE:** Academic titles (e.g., `Dr.`, `Dr.-Ing.`, `Ing.`) as they are part of the person's formal name.
    *   **INCLUDE:** The word "Firma" if it appears as part of the company name (e.g., `Firma C. KESSELER`).
    *   **CRITICAL - EXCLUDE OCCUPATIONS:** NEVER include occupational descriptions or job titles, such as:
        *   `Hofschlächter`, `Professor`, `Ingenieur`, `Kaufmann`, `Fabrikant`, `Mechaniker`
        *   `Nachfolger`, `Marine-Lieutenant`, or any other job titles
    *   **CRITICAL - EXCLUDE LOCATIONS:** NEVER include location information in the name field. For multiple patent holders, extract ONLY the names and separate them with `und` or `,` as they appear in the original text.
*   **CRITICAL:** This field is for the patent holder ONLY. Do NOT extract the name of the representative (`Vertreter`). The representative's name, if present, appears later in the text and must be ignored for this field.

**3. `location` (String)**
*   **Position:** This is the location of the patent holder(s), and it almost always follows the holder's `name` directly. It is often introduced by a preposition like `in` or `auf`.
*   **Rule:** Extract the entire location phrase associated with the patent holder(s).
    *   Include all details: city, state, country (e.g., `Kladno, Böhmen`, `Chicago, V. St. A.`).
    *   **CRITICAL - MULTIPLE PATENT HOLDERS:** When there are multiple patent holders, each with their own location, concatenate ALL locations into a single string, separated by `und` or `,` as they appear in the original text.
    *   **Example:** For `RUHM, H., in Sulko Zechc bei Pilsen und L. WOLF in Prag-Karolinenthal`, the location should be `Sulko Zechc bei Pilsen und Prag-Karolinenthal`.
*   **CRITICAL:** Do NOT extract the location of the representative (`Vertreter`). The correct location is the one directly linked to the `name` you extracted in the previous step.

**4. `description` (String)**
*   **Position:** This is the patent's subject matter. It is the text located *after* the holder's name/location and *before* the final date.
*   **Rule:** Extract the core descriptive text. This can be as short as a single word. You must apply the following two exclusion rules:
    *   **Exclusion 1: Representative Information:** You MUST completely exclude any block of text detailing a representative. This block often starts with `Vertreter:` and includes their name, title, and location.
    *   **Exclusion 2: Addendum Notes:** You MUST completely exclude any parenthetical notes about patent addendums after or at the end of the patent description. These notes look like `(Zusatz zum Patent 12345.)` or `(II. Zusatz zum Patent 17502.)` and are usually at the end or after the patent description.
*   After applying these exclusions, extract the remaining text for the description.

**5. `date` (String)**
*   **Position:** The date is located near the end of the entry.
*   **Rule:** Extract the date string *exactly* as it appears. If multiple dates are present in the entry, the correct one is always the date referring to the German registration.

---

**Example Input Entry:**
35 321. — KARLIK, JOH., in Kladno, Böhmen; Vertreter: C. FEHLERT & G. LOUBIER, Firma: C. KESSELER in Berlin SW., Anhaltstr. 6. Wipper mit verschiedenen, sich während einer Umdrehung ändernden Umfangsgeschwindigkeiten. 5. November 1885.

**Correct JSON Output:**
{
  "patent_id": "35321",
  "name": "KARLIK, JOH.",
  "location": "Kladno, Böhmen",
  "description": "Wipper mit verschiedenen, sich während einer Umdrehung ändernden Umfangsgeschwindigkeiten.",
  "date": "5. November 1885"
}

**Second Example Input Entry:**
15948. — „VIEILLE MONTAGNE" AKTIEN-GESELLSCHAFT FÜR BERGBAU- UND ZINKHÜTTENBETRIEB in Altenberge. Neuerungen an ringförmigen Setzmaschinen (Zusatz zu P. R. 15224). 25. März 1881. — A 789.

**Correct JSON Output:**
{
  "patent_id": "15948",
  "name": "\"VIEILLE MONTAGNE\" AKTIEN-GESELLSCHAFT FÜR BERGBAU- UND ZINKHÜTTENBETRIEB",
  "location": "Altenberge",
  "description": "Neuerungen an ringförmigen Setzmaschinen",
  "date": "25. März 1881"
}

**Third Example Input Entry (Multiple Patent Holders and Locations):**
15203. — RUHM, H., in Sulko Zechc bei Pilsen und L. WOLF in Prag-Karolinenthal; Vertreter: C. KESSELER in Berlin W., Mohrenstr. 63 I. Endloser Planherd. 1. April 1881. — A 609.

**Correct JSON Output:**
{
  "patent_id": "15203",
  "name": "RUHM, H. und L. WOLF",
  "location": "Sulko Zechc bei Pilsen und Prag-Karolinenthal",
  "description": "Endloser Planherd.",
  "date": "1. April 1881"
}

**The entry from which you should extract information:**
\end{lstlisting}
\end{tcolorbox}

\begin{minipage}{\linewidth}
    \footnotesize
    \justifying
    \begin{spacing}{1}
    \noindent\textit{Notes}: This prompt was used to extract the five desired variables. A repaired patent entry was appended to the bottom of this prompt and then sent to Gemini-2.5-Flash-Lite. The LLM returned a JSON object with keys \textit{patent\_id}, \textit{name}, \textit{location}, \textit{description}, and \textit{date}. The variables were then appended to the respective patent entry in the dataset. This was done indepdently for all patent entries.
    \end{spacing}
\end{minipage}

%% file: 10_appendix_b.tex
\newpage
\section*{Appendix B: Prompts for Special Volumes with Different Layout}
\label{sec:appendix_b}
\setappendixfigures{B}

\begin{center}
    \captionsetup{hypcap=false}
    \captionof{figure}{\textsc{Special Volumes with Different Layout: Patent Entry Extraction Prompt}}
    \label{fig:prompt-setup}
\end{center}

\vspace{-1.75em}

\begin{tcolorbox}[
    breakable,
    enhanced,
    colback=gray!5!white,
    colframe=black,
    coltitle=white,
    fonttitle=\bfseries,
    boxrule=0.5pt,
    arc=0mm,
    top=0mm, bottom=0mm, left=0mm, right=0mm,
    boxsep=1mm
]
\begin{lstlisting}[
    basicstyle=\scriptsize\ttfamily,
    breaklines=true,
    breakatwhitespace=true,
    columns=fullflexible,
    keepspaces=true,
    aboveskip=5pt, belowskip=5pt,
    % --- CHARACTER MAPPING (Kept active for safety) ---
    literate=
        {ö}{{\"o}}1
        {ä}{{\"a}}1
        {ü}{{\"u}}1
        {Ö}{{\"O}}1
        {Ä}{{\"A}}1
        {Ü}{{\"U}}1
        {ß}{{\ss}}1
        {—}{{---}}1   % Em-dash
        {–}{{--}}1    % En-dash
        {−}{{--}}1    % Minus sign
        {†}{{\dag}}1  % Dagger
        {ſ}{{s}}1     % Long s
        {А}{{A}}1     % Cyrillic A
]
You are a specialist data extraction AI. Your task is to process a single page from a German Imperial Patent Office register and extract the specified contents into a perfectly structured, flat JSON array. You must act with the precision of a world-class quantitative historian, ensuring absolute accuracy and adherence to the following rules. **HISTORICAL ACCURACY IS PARAMOUNT - extract everything exactly as written, preserving all abbreviations, spellings, and punctuation exactly as they appear in the original text.**

## I. Core Principles & Scope

### Extraction Area
You will **ONLY** extract information from the two main content columns on the page.

### IGNORE Absolutely:
*   **Running Page Headers:** The text at the very top of the page (e.g., "Kl. 18. Nr. 55711." or "Kl. 19. Nr. 57 185." at the top-left and top-right).
*   **Page Numbers:** Any numbers located in the center of the running header.
*   **Footer Information:** Any text at the bottom of the page that begins with an asterisk (*), is a correction ("Berichtigung"), or contains standalone numbers (e.g., "2*", "1917") that are clearly not part of the patent entries above.

### Reading & Processing Order
Your output must strictly follow this sequence:
1.  First, process the **LEFT** content column from top to bottom.
2.  Second, process the **RIGHT** content column from top to bottom.

The final JSON array must reflect this sequential order of extracted items.

## II. Extraction Rules: Identifying & Formatting Content

For each block of text you encounter in the two main content columns, you must classify it as one of the following and format it accordingly.

### A. Category Headings

*   **Identification:** A category heading is a line that signals a new patent class.
    *   A main class heading always starts with "Klasse", followed by a number and a title (e.g., "Klasse 19. Eisenbahn- und Brückenbau.").
*   **Extraction Rule:** From the identified heading, you must extract **ONLY** the class number and nothing else.
    *   If the heading is "Klasse 19. Eisenbahn- und Brückenbau", you will extract "19".
    *   Clean the extracted code by removing any trailing periods or whitespace.
*   **JSON Output:** The extracted class code must be placed in an object with the key "category".
    > **Example:** {"category": "3"}
    > **Example:** {"category": "19"}

### B. Patent Entries

*   **Identification of a Standard Entry:** A standard, complete patent entry is a paragraph that:
    *   Contains descriptive text about the patent holder, location, title, description, etc.
    *   Always ends with a patent registration number in the format "P. R. XXXX." (where XXXX is a number between 1 and 9999, e.g., "P. R. 6201.").

*   **Extraction Rule for ALL Entries (Standard and Truncated):**
    *   Capture the text of the entry verbatim, preserving all original characters, spelling, and punctuation.
    *   **Line Break Handling:**
        *   Join words that are hyphenated across a newline (e.g., Hindersin-\nstraſse becomes Hindersinstraſse).
        *   Replace all other newlines within the entry's paragraph with a single space to form one continuous line of text.

*   **CRITICAL EXCEPTION: Truncated Entries**
    The single-page context means you will encounter incomplete entries. You **MUST** extract these as they appear.
    *   **Entry Truncated at the Top-Left:** If a column begins mid-paragraph (continuing from a previous page), extract that partial entry as-is.
    *   **Entry Truncated at the Bottom-Right:** If an entry is cut off at the bottom of a column (continuing to the next page), extract that partial entry as-is.
    *   **Entry Spanning Columns on the SAME Page:** If an entry starts at the bottom of the left column and its text continues at the top of the right column, you must create **TWO SEPARATE** entries in the JSON array. The first object will contain the part from the left column, and the second object will contain the part from the right column.

*   **JSON Output:** The formatted text of the entry must be placed in an object with the key "entry".
    > **Example:** {"entry": "COOMES, M, F., Arzt, und A. W. HYDE, in Louisville, 6281/2 Fourth Street bezw. 828 Second Street, Grafschaft Jefferson, Staat Kentucky, V. St. A.; Vertreter: C. PIEPER in Berlin N.W., Hindersinstraſse 3. Verfahren zur Bereitung von Stahl. 26. Februar 1890. P. R. 6201."}

### C. Content to Explicitly IGNORE
*   **Lists of Patent Numbers:** After a category heading, you will often see a block of text that is just a list of patent numbers and codes (e.g., 54571 A 3. — 54814 A 77. — 54881 A 37. — 54977 A 54.). **DO NOT** extract these lists that are just consisting of patent numbers and patent codes. These invalid patent entries only occur directly below category headings. **These lists can span multiple lines and form large blocks of text - ignore the entire block.** Dates in a new line below a valid patent entry belong to the valid patent entry.
*   **Reference Information ("Vgl."):** Any text that begins with "Vgl." (e.g., "Vgl. Kl. 12 P. R. 21188. – Kl. 45 P. R. 17470, 20344.") that appears between patent entries and class headings. **DO NOT** extract this cross-reference information.

## III. JSON Output Specification (Strict)

Your entire output must be a single, syntactically perfect JSON array. Any deviation will result in failure.

*   **Array Structure:** The output must begin with [ and end with ].
*   **Object Structure:** Each element in the array is an object {}.
*   **Key-Value Pairs:** Each object must contain exactly **one** key-value pair. The key must be either "category" or "entry".
*   **Quotation:** **ALL** keys **MUST** be enclosed in double quotes (e.g., "entry"). All string values must be enclosed in double quotes.
*   **Delimiters:** Objects must be separated by a comma ,. The last object in the array must **NOT** have a trailing comma.
*   **Escaping:** Within a string value, any literal double quote " must be escaped as \".

---

**Example of Perfect Output Structure:**

[
  {
    "category": "18"
  },
  {
    "entry": "COOMES, M, F., Arzt, und A. W. HYDE in Louisville, 6281/2 Fourth Street bezw. 828 Second Street, Grafschaft Jefferson, Staat Kentucky, V. St. A.; Vertreter: C. PIEPER in Berlin N.W., Hindersinstraſse 3. Verfahren zur Bereitung von Stahl. 26. Februar 1890. P. R. 6201."
  },
  {
    "entry": "VERSEN, B., in Dortmund, Heiligerweg 8. Verfahren und Vorrichtung zur Herstellung von Bessemer-Birnen-Böden. (Zusatz zum Patente Nr. 30634.) 15. Mai 1890. P. R. 6202."
  },
  {
    "entry": "ADAMS, CH., in Pittsburgh, 110 Diamond Street, Pennsylvania, V. St. A.; Vertreter: F. EDMUND THODE & KNOOP in Dresden, Amalienstraſse 5. Unmittelbare Darstellung von Eisen aus seinen Erzen. 9. Juli 1890. P. R. 6203."
  },
  {
    "entry": "HERBERTZ, F. A., in Köln a. Rh. Schmelzofen mit Dampfstrahl. 12. August 1890. P. R. 6204."
  },
  {
    "category": "19"
  },
  {
    "entry": "FÜRSTENHEIM, C., in Berlin C., Jerusalemerstraſse 15. Zusammenklappbares Büstengestell. 5. Juni 1890. P. R. 6205."
  }
]

**Final Instruction:**  Adhere to all rules. Your output must be **ONLY the JSON array** and nothing else. Before finishing, double-check your output to ensure it is perfectly valid JSON according to the specifications above.
\end{lstlisting}
\end{tcolorbox}

\begin{minipage}{\linewidth}
    \footnotesize
    \justifying
    \begin{spacing}{1}
    \noindent\textit{Notes}: The prompt used to extract patent entries and technological classes from images in volumes 1877--8 and 1879, which have a different layout from the other volumes. Please see the notes in Figure~\ref{fig:01-prompt} for more information.
    \end{spacing}
\end{minipage}

\begin{center}
    \captionsetup{hypcap=false}
    \captionof{figure}{\textsc{Special Volumes with Different Layout: Reparation Prompt}}
    \label{fig:prompt-validation}
\end{center}

\vspace{-1.5em}

\begin{tcolorbox}[
    breakable,
    enhanced,
    colback=gray!5!white,
    colframe=black,
    coltitle=white,
    fonttitle=\bfseries,
    boxrule=0.5pt,
    arc=0mm,
    top=0mm, bottom=0mm, left=0mm, right=0mm,
    boxsep=1mm
]
\begin{lstlisting}[
    basicstyle=\scriptsize\ttfamily,
    breaklines=true,
    breakatwhitespace=true,
    columns=fullflexible,
    keepspaces=true,
    aboveskip=5pt, belowskip=5pt,
    % --- CHARACTER MAPPING ---
    literate=
        {ö}{{\"o}}1
        {ä}{{\"a}}1
        {ü}{{\"u}}1
        {Ö}{{\"O}}1
        {Ä}{{\"A}}1
        {Ü}{{\"U}}1
        {ß}{{\ss}}1
        {—}{{---}}1   % Em-dash
        {–}{{--}}1    % En-dash
        {−}{{--}}1    % Minus sign
        {†}{{\dag}}1  % Dagger
        {ſ}{{s}}1     % Long s
        {„}{{,,}}1    % German opening quote
        {“}{{''}}1    % German closing quote
]
You are a data validator. Your task is to classify a German patent entry as either complete (`1`) or truncated (`0`).

A **complete (valid)** entry must end with a date followed by a patent registration number. Due to OCR errors, the registration format may appear in many variations including "P. R. XXXX.", "I. R. XXXX.", "R. XXXX.", "R. R. XXXX.", "P. R. Nr. XXXX", "P. R. Nr. XXXX.", etc. (where XXXX is a number between 1 and 9999). The key is that it ends with a date and some form of registration number. An entry is complete even if it's truncated at the beginning, as long as it ends properly.

**Examples of COMPLETE entries (should be marked as `1`):**
- `HAHLWEG, C. Werkzeug zur Herstellung von Steinfassungen für Taschenuhren. 22. Juli 1877. P. R. 80.`
- `LÖWIG, G., & LÖWIG, Fr. Verfahren zur Darstellung von Aetzalkalien und Thonerdepräparaten. 3. Juli 1877. P. R. 93.`
- `MORGAN RICHARDS, J. Herstellung durchbohrter Pillen und der zu ihrer Anfertigung nöthigen Maschine. 3. August 1877. P. R. 134.`
- `anilin und anderen tertiären aromatischen Monaminen. 15. Dezember 1877. P. R. 1886.` (truncated at beginning but complete)
- `SCHRÖDER, C. Windmühlenflügel mit durch Federn geöffneten, durch den Winddruck geschlossenen Klappen. 13. Januar 1878. I. R. 1843.` (OCR error: I. R. instead of P. R.)
- `BERNHARDI SOHN, Dr., DRAENERT, G. E. JACKSON'sche Wendevorrichtung an horizontalen Windrädern mit Kettenbetrieb. 13. November 1877. R. 1229.` (OCR error: R. instead of P. R.)
- `SCHNEIDER & JACQUED. Turbine mit eingesenkten Zwischenschaufeln. 4. September 1877. R. 547.` (OCR error: R. instead of P. R.)
- `FISCHER, G. A. Niederschraubhahn mit und ohne Entleerungsventil, mittelst dessen eine unter Druck stehende Dampf- oder Wasserleitung angebohrt werden kann. 20. Oktober 1877. R. R. 874.` (OCR error: R. R. instead of P. R.)
- `GROSSCHOPFF, DR. C. Selbstthätiger Desinfector für Aborte. 31. März 1878. P. R. Nr. 3197` (OCR variation: P. R. Nr. without final period)

An entry is **truncated (invalid)** if it is missing the date or patent registration number at the very end.

**Examples of TRUNCATED entries (should be marked as `0`):**
- `BARDIN, G. Verfahren, die natürliche Feder zu verzwirnen und die Verwendung solcher Feder-Che`
- `EHESTÄDT und ROBERT. Beweglicher Arm an Beleuchtungsapparaten aus in Form eines Parallelo-`

**CRITICAL:** When in doubt, mark as `1` (complete). Only mark as `0` if the entry is clearly truncated and missing both date and registration number at the end.

Respond with a single character only:
- `1` if the entry is complete (ends with date and any form of registration number, regardless of OCR variations).
- `0` if the entry is truncated/invalid (missing date and registration number at the end).

Do not provide any explanation or additional text.
\end{lstlisting}
\end{tcolorbox}

\begin{minipage}{\linewidth}
    \footnotesize
    \justifying
    \begin{spacing}{1}
    \noindent\textit{Notes}: The prompt used to repair patent entries from volumes 1877--8 and 1879, which have a different layout from the other volumes. Please see the notes in Figure~\ref{fig:02-prompt} for more information.
    \end{spacing}
\end{minipage}

\begin{center}
    \captionsetup{hypcap=false}
    \captionof{figure}{\textsc{Special Volumes with Different Layout: Variable Extraction Prompt}}
    \label{fig:prompt-extraction-type-b}
\end{center}

\vspace{-1.5em}

\begin{tcolorbox}[
    breakable,
    enhanced,
    colback=gray!5!white,
    colframe=black,
    coltitle=white,
    fonttitle=\bfseries,
    boxrule=0.5pt,
    arc=0mm,
    top=0mm, bottom=0mm, left=0mm, right=0mm,
    boxsep=1mm
]
\begin{lstlisting}[
    basicstyle=\scriptsize\ttfamily,
    breaklines=true,
    breakatwhitespace=true,
    columns=fullflexible,
    keepspaces=true,
    aboveskip=5pt, belowskip=5pt,
    % --- CHARACTER MAPPING ---
    literate=
        {ö}{{\"o}}1
        {ä}{{\"a}}1
        {ü}{{\"u}}1
        {Ö}{{\"O}}1
        {Ä}{{\"A}}1
        {Ü}{{\"U}}1
        {ß}{{\ss}}1
        {—}{{---}}1   % Em-dash
        {–}{{--}}1    % En-dash
        {−}{{--}}1    % Minus sign
        {†}{{\dag}}1  % Dagger
        {ſ}{{s}}1     % Long s
        {„}{{,,}}1    % German opening quote
        {“}{{''}}1    % German closing quote
]
You will be provided with a single text entry from Germany's Imperial Patent Office. Your task is to function as a highly precise information extraction engine. You must carefully analyze the entry and extract specific fields, formatting the output as a single, valid JSON object.

**Core Principles:**
*   **Exactness:** All extracted text must be an *exact, character-for-character copy* from the source text. Do not add, remove, or alter any characters, including historical German characters (e.g., ſ).
*   **Structure:** The patent entries almost always follow a specific order: `patent_id` -> `name(s)` -> `location(es)` -> (optional `Vertreter` info) -> `description` -> (optional addendum note) -> `date` -> (optional addendum note) -> `P. R. XXXX.` (registration number). Use this predictable structure to guide your extraction.
*   **Output Format:** The output MUST be a single JSON object with the keys: `patent_id`, `name`, `location`, `description`, and `date`. If a value cannot be reliably found or identified for a field, use the string "NaN".

---

**Detailed Extraction Rules:**

**1. `patent_id` (String)**
*   **Location:** The patent ID is the numerical identifier at the end of the entry in the format "P. R. XXXX.".
*   **Format:** It is a number between 1 and 9999.
*   **Rule:** Extract ONLY the numeric digits from the "P. R. XXXX." format at the end of the entry.
*   **Example:** For `... 5. November 1885. P. R. 6201.`, you must extract `"6201"`.

**2. `name` (String)**
*   **Location:** This is the name of the patent holder(s), which appears at the beginning of the entry.
*   **Rule:** Extract the full name(s) exactly as written.
    *   Include corporate forms (e.g., `AKTIEN-GESELLSCHAFT FÜR BERGBAU- UND ZINKHÜTTENBETRIEB`).
    *   Include academic or professional titles (e.g., `Dr.-Ing.`).
    *   Include all parts of compound names (e.g., `SCHIFFER & KIRCHER`).
    *   **CRITICAL:** Do NOT extract occupations or job descriptions that may appear with names. Extract only the actual names and titles.
*   **CRITICAL:** This field is for the patent holder ONLY. Do NOT extract the name of the representative (`Vertreter`). The representative's name, if present, appears later in the text and must be ignored for this field.

**3. `location` (String)**
*   **Location:** This is the location of the patent holder, and it almost always follows the holder's `name` directly. It is often introduced by a preposition like `in` or `auf`.
*   **Rule:** Extract the entire location phrase associated with the patent holder(s).
    *   Include all details: city, state, country (e.g., `Kladno, Böhmen`, `Chicago, V. St. A.`).
*   **CRITICAL:** Do NOT extract the location of the representative (`Vertreter`). The correct location is the one directly linked to the `name` you extracted in the previous step.

**4. `description` (String)**
*   **Location:** This is the patent's subject matter. It is the text located *after* the holder's name/location and *before* the final date.
*   **Rule:** Extract the core descriptive text. This can be as short as a single word. You must apply the following two exclusion rules:
    *   **Exclusion 1: Representative Information:** You MUST completely exclude any block of text detailing a representative. This block often starts with `Vertreter:` and includes their name, title, and location.
    *   **Exclusion 2: Addendum Notes:** You MUST completely exclude any parenthetical notes about patent addendums. These notes look like `(Zusatz zum Patent 12345.)` or `(II. Zusatz zum Patent 17502.)` and are usually found at the end of the description, just before the date.
*   After applying these exclusions, extract the remaining text for the description.

**5. `date` (String)**
*   **Location:** The date is located near the end of the entry, before the final "P. R. XXXX." registration number.
*   **Rule:** Extract the date string *exactly* as it appears. If multiple dates are present in the entry, the correct one is always the date referring to the German registration.

---

**Example Input Entry:**
KARLIK, JOH., in Kladno, Böhmen; Vertreter: C. FEHLERT & G. LOUBIER, in Firma: C. KESSELER in Berlin SW., Anhaltstrafse 6. Wipper mit verschiedenen, sich während einer Umdrehung ändernden Umfangsgeschwindigkeiten. 5. November 1885. P. R. 6201.

**Correct JSON Output:**
{
  "patent_id": "6201",
  "name": "KARLIK, JOH.",
  "location": "Kladno, Böhmen",
  "description": "Wipper mit verschiedenen, sich während einer Umdrehung ändernden Umfangsgeschwindigkeiten.",
  "date": "5. November 1885."
}

**The entry from which you should extract information:**
\end{lstlisting}
\end{tcolorbox}

\begin{minipage}{\linewidth}
    \footnotesize
    \justifying
    \begin{spacing}{1}
    \noindent\textit{Notes}: The prompt used to extract variables from patent entries stemming from volumes 1877--8 and 1879, which have a different layout from the other volumes. Please see the notes in Figure~\ref{fig:03-prompt} for more information.
    \end{spacing}
\end{minipage}